\documentclass[a4paper, 11pt, oneside]{article}
\pdfoutput=1
\usepackage[a4paper,inner=2cm,outer=2cm,top=3cm,bottom=2.5cm]{geometry}

\usepackage[tbtags]{amsmath}
\usepackage{amssymb}
\usepackage{amsthm}
\usepackage{dsfont}
\usepackage{slashed}
\usepackage{mathrsfs}
\usepackage[tbtags]{mathtools}
\usepackage{color}

\usepackage[export]{adjustbox}

\usepackage[english]{babel}

\usepackage[utf8x]{inputenc}

\usepackage{graphicx,epsfig}

\usepackage{setspace}

\usepackage{booktabs}

\usepackage{float}

\usepackage{dcolumn}

\usepackage{nextpage}

\usepackage{newclude}

\usepackage[small,bf]{caption}
\usepackage{subcaption}

\usepackage{url}

\usepackage[colorlinks=true,
	linkcolor=blue,
	citecolor=blue,
	urlcolor=blue,
	filecolor=black
]{hyperref}

\usepackage[T1]{fontenc}

\usepackage{cancel}

\usepackage{tabularx}

\usepackage{pifont}

\usepackage{etex}

\usepackage{appendix}

\usepackage[numbers,sort&compress]{natbib}

\usepackage{placeins}

\usepackage{ragged2e}

\usepackage{ifmtarg}

\usepackage{tikz}
\usetikzlibrary{shapes,arrows}

\usepackage{bchart}

\renewcommand{\O}{\mathcal{O}}

\renewcommand{\Im}{\text{Im}\,}

\newcommand{\<}{\langle}
\renewcommand{\>}{\rangle}

\newcommand{\mpi}{M_{\pi}}

\newcommand{\BR}{\text{BR}}

\newcommand{\beq}{\begin{equation}}
\newcommand{\eeq}{\end{equation}}
\newcommand{\GeV}{\,\text{GeV}}

\newcommand{\remark}[1]{}

\newcommand{\mytag}{\\[-\baselineskip] \stepcounter{equation}\tag{\theequation}}

\makeatletter
\newcommand{\Cr}[2]{\@ifmtarg{#2}{\mathcal{C}_{#1}}{\mathcal{C}_{#1}\big[#2\big]}}
\makeatother

\newcommand{\nn}{\nonumber\\}

\newcolumntype{.}{D{.}{.}{2} } % decimal columns
\newcolumntype{d}{D{.}{.}{2.2} }
\newcolumntype{L}[1]{>{\RaggedRight\hspace{0pt}}p{#1}}
\newcolumntype{R}[1]{>{\RaggedLeft\hspace{0pt}}p{#1}}

\makeatletter
  \def\my@tag@font{\normalsize}
  \def\maketag@@@#1{\hbox{\m@th\normalfont\my@tag@font#1}}
  \let\amsmath@eqref\eqref
  \renewcommand\eqref[1]{{\let\my@tag@font\relax\amsmath@eqref{#1}}}
\makeatother

\makeatletter
\renewcommand\paragraph{\@startsection{paragraph}{4}{\z@}%
  {-3.25ex \@plus -1ex \@minus -0.2ex}%
  {0.01pt}%
  {\bfseries}%
}
\makeatother

\makeatletter
\def\@xfootnote[#1]{%
  \protected@xdef\@thefnmark{#1}%
  \@footnotemark\@footnotetext}
\makeatother

\usepackage{abstract}

\allowdisplaybreaks[1]

\begin{document}

\mbox{}

\vspace{-1.75cm}
\hfill{}\begin{minipage}[t][0cm][t]{5cm}
\raggedleft
\footnotesize
INT-PUB-19-023 
\end{minipage}
\vspace{1.25cm}

\bigskip

\begin{center}
{\LARGE{\bf \boldmath Dispersion relations for $\gamma^*\gamma^*\to\pi\pi$: helicity\\[2mm] amplitudes, subtractions, and anomalous thresholds}}

\vspace{0.5cm}

Martin Hoferichter${}^{a}$, Peter Stoffer${}^{b}$

\vspace{1em}

\begin{center}
\it
\mbox{} \\
${}^a$Institute for Nuclear Theory, University of Washington, Seattle, WA 98195-1550, USA \\
\mbox{}\\
${}^b$Department of Physics, University of California at San Diego, La Jolla, CA 92093, USA
\end{center} 

\end{center}

\vspace{3em}

\hrule

\begin{abstract}
  We present a comprehensive analysis of the dispersion relations for the doubly-virtual process $\gamma^*\gamma^*\to\pi\pi$. Starting from the 
  Bardeen--Tung--Tarrach amplitudes, we first derive the kernel functions that define the system of Roy--Steiner equations for the partial-wave 
  helicity amplitudes. We then formulate the solution of these partial-wave dispersion relations in terms of Omn\`es functions, with special attention
  paid to the role of subtraction constants as critical for the application to hadronic light-by-light scattering.
  In particular, we explain for the first time why for some amplitudes the standard Muskhelishvili--Omn\`es solution applies, 
  while for others a modified approach based on their left-hand cut is required unless subtractions are introduced. 
  In the doubly-virtual case, the analytic structure of the vector-resonance partial waves then gives rise to anomalous thresholds, even for space-like virtualities.
  We develop a strategy to account for these effects in the numerical solution, illustrated 
  in terms of the $D$-waves in $\gamma^*\gamma^*\to\pi\pi$, which allows us to predict the doubly-virtual responses of the $f_2(1270)$ resonance. 
  In general, our results form the basis for the incorporation of two-meson intermediate states into hadronic light-by-light scattering beyond the $S$-wave contribution. 
\end{abstract}

\hrule

%\newpage

\setcounter{tocdepth}{3}
\tableofcontents

\numberwithin{equation}{section}

	% !TEX root = ../DV_Paper.tex

\section{Introduction}

Apart from the two-photon decay of the neutral pion, the reaction $\gamma\gamma\to\pi\pi$ constitutes the simplest process that gives access to 
the electromagnetic properties of the pion, most notably its dipole polarizabilities~\cite{Holstein:2013kia}.
Experimentally, most information on the scattering process comes from $e^+e^-$ colliders via the reaction 
$e^+e^-\to e^+e^-\pi\pi$~\cite{Marsiske:1990hx,Boyer:1990vu,Behrend:1992hy,Mori:2007bu,Uehara:2008ep,Uehara:2009cka},
while the kinematics relevant for the extraction of the polarizabilities is more directly probed
in the Primakoff process, where an incident pion scatters of the Coulomb field of a heavy nucleus
and produces a final-state photon--pion pair~\cite{Pirmakoff:1951pj,Antipov:1982kz,Adolph:2014kgj}.
The measurement of the polarizabilities, as well as the energy dependence of the $\gamma\gamma\to\pi\pi$ cross section, 
also provides a key test of chiral perturbation theory (ChPT)~\cite{Weinberg:1966kf,Weinberg:1978kz,Gasser:1983yg,Gasser:1984gg}, 
not only because it is the simplest electromagnetic scattering process involving hadrons, but also 
due to the sensitivity to chiral loop corrections, see~\cite{Bijnens:1987dc,Donoghue:1988eea} 
and~\cite{Bellucci:1994eb,Burgi:1996mm,Burgi:1996qi,Gasser:2005ud,Gasser:2006qa} for the one- and two-loop calculation, respectively.
While an extraction of the charged-pion polarizability via radiative pion production off the nucleon~\cite{Ahrens:2004mg} had
been interpreted as a potential tension with ChPT~\cite{Gasser:2006qa}---despite the model-dependence from the extrapolation to the pion pole---the
most recent Primakoff measurement~\cite{Adolph:2014kgj} confirmed the chiral prediction. 
In addition to a future update from COMPASS~\cite{Friedrich:2019pui}, further low-energy measurements that would entail 
additional information on the charged-pion polarizabilities are planned at Hall D at Jefferson Lab 
via the Primakoff process with an incident photon~\cite{Lawrence:2013asa}.

To extend the description of $\gamma\gamma\to\pi\pi$ beyond the low-energy region, dispersion relations (DRs) have been widely applied in the literature~\cite{Gourdin:1960,Babelon:1976ww,Morgan:1987gv,Donoghue:1993kw,Drechsel:1999rf,Filkov:2005suj,Pennington:2008xd,Oller:2007sh,Oller:2008kf,Mao:2009cc,GarciaMartin:2010cw,Hoferichter:2011wk,Moussallam:2011zg,Dai:2014zta}, most importantly to include the strong $\pi\pi$ rescattering in the $S$-wave. More recently, this method has been extended to a single off-shell photon~\cite{Moussallam:2013una,Danilkin:2018qfn}, as well as the doubly-virtual case~\cite{Colangelo:2015ama,Colangelo:2017qdm,Colangelo:2017fiz}, with 
numerical results provided for the $S$-wave contribution. 
In this paper, we address, comprehensively, the general case in which both photons are virtual, as required as input for a dispersive approach to hadronic light-by-light (HLbL) scattering in the anomalous magnetic moment of the muon $g-2$~\cite{Hoferichter:2013ama,Colangelo:2014dfa,Colangelo:2014pva,Colangelo:2015ama,Colangelo:2017qdm,Colangelo:2017fiz,Hoferichter:2018dmo,Hoferichter:2018kwz}. 
In particular, we consider several technical challenges that appear in the contribution of two-meson intermediate states beyond the $S$-waves.

First of all, the ChPT amplitudes for the doubly- (or even singly-) virtual case have only been worked out at one-loop order~\cite{Colangelo:2014dfa}. Since the one-loop contribution does not display any angular dependence except for the charged-pion Born terms, this implies that chiral predictions for $D$- and higher partial waves are not available. Second, it was shown in~\cite{GarciaMartin:2010cw}
that an adequate description of the $D$-waves requires the inclusion of vector mesons in the left-hand cut (LHC) of the $\gamma\gamma\to\pi\pi$ amplitudes, most efficiently in terms
of the LHC of the partial waves. This strategy extends the standard Muskhelishvili--Omn\`es (MO) solution~\cite{Omnes:1958hv,Muskhelishvili:1953}, and, as we will show here, its necessity is related to 
the high-energy behavior of the vector-meson partial waves and thus
potential subtractions in the MO solution.
Third, the derivation of partial-wave DRs for the helicity amplitudes has to be based on scalar functions that avoid kinematic singularities and zeros~\cite{Bardeen:1969aw,Tarrach:1975tu}. 
Based on the corresponding set of amplitudes from~\cite{Colangelo:2015ama} we explicitly write down the kernel functions that couple the various partial waves and 
perform a basis change that diagonalizes their MO solution. 
Finally, we observe that if vector resonances are to be included in the MO solution in terms of the LHC also in the doubly-virtual case, the analytic structure of these amplitude complicates the 
implementation for sufficiently large (space-like) virtualities. While the occurrence of anomalous thresholds~\cite{Mandelstam:1960zz} is expected in the time-like regime~\cite{Lucha:2006vc,Hoferichter:2013ama,Colangelo:2015ama}, the analytic structure of the resonance LHCs is sufficiently complicated that even in the space-like case
a deformation of the integration contour becomes unavoidable.

We first recall the definition of helicity amplitudes, partial waves, and Bardeen--Tung--Tarrach (BTT) invariant functions in Sect.~\ref{sec:amplitudes}, based on which we then derive all relevant kernel functions that define the full system of Roy--Steiner (RS) equations for the partial-wave helicity amplitudes.
In Sect.~\ref{sec:MO_solution} we then write down the MO solution of these equations, and discuss in detail the role of subtraction constants as well as the analytic structure of the 
resonance partial waves. Some numerical results will be presented in Sect.~\ref{sec:numerics}, before we conclude in Sect.~\ref{sec:conlusions} and 
comment on the implications of our results for the application to HLbL scattering.

	% !TEX root = ../DV_Paper.tex

\section{Helicity amplitudes and Roy--Steiner equations}
\label{sec:amplitudes}

\begin{figure}[t]
	\centering
	\includegraphics[width=4cm]{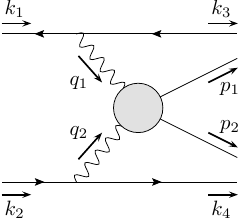}
	\caption{$\gamma^*\gamma^*\to\pi\pi$ as a sub-process of $e^+e^-\to e^+e^-\pi\pi$~\cite{Colangelo:2015ama}.}
	\label{img:eetopipi}
\end{figure}

\subsection{Helicity amplitudes}
\label{sec:helicity_amplitudes}

We largely follow the conventions of~\cite{Colangelo:2015ama}, but for completeness repeat the basic definitions. The process $\gamma^*\gamma^*\to\pi\pi$ is strictly speaking not
observable, but derived from processes with well-defined asymptotic states under certain assumptions. We start from
\begin{align}
	e^+(k_1) e^-(k_2) \to e^+(k_3) e^-(k_4) \gamma^*(q_1) \gamma^*(q_2) \to e^+(k_3) e^-(k_4) \pi^a(p_1) \pi^b(p_2),
\end{align}
shown in Fig.~\ref{img:eetopipi}, with isospin labels $a$ and $b$ for the pion states and momenta as indicated. 
At $\O(e^4)$, the amplitude for this process is given by
\begin{align}
\label{T_eeeepipi}
	\begin{split}
		i \mathcal{T} &= \bar v(k_1) (-i e \gamma_\alpha) v(k_3) \bar u(k_4) (-i e \gamma_\beta) u(k_2) \\
			& \quad \times \frac{-i}{q_1^2} \left( g^{\alpha\mu} - (1-\xi) \frac{q_1^\alpha q_1^\mu}{q_1^2} \right) \frac{-i}{q_2^2} \left( g^{\beta\nu} - (1-\xi) \frac{q_2^\beta q_2^\nu}{q_2^2} \right) i e^2 W_{\mu\nu}^{ab}(p_1,p_2,q_1) ,
	\end{split}
\end{align}
where $\xi$ is a gauge parameter for the photon propagators and the tensor
\begin{align}
	W^{\mu\nu}_{ab}(p_1,p_2,q_1) = i \int d^4x \, e^{-i q_1 \cdot x} \<\pi^a(p_1) \pi^b(p_2) | T \{ j_\mathrm{em}^\mu(x) j_\mathrm{em}^\nu(0) \} | 0 \>
\end{align}
is defined in pure QCD.
The contraction of this tensor with appropriate polarization vectors is then identified as an amplitude for the off-shell process
\begin{align}
	\gamma^*(q_1,\lambda_1) \gamma^*(q_2,\lambda_2) \to \pi^a(p_1) \pi^b(p_2) ,
\end{align}
where $\lambda_{1,2}$ denote the helicities of the photons. The connected part is given by
\begin{align}
	\< \pi^a(p_1) & \pi^b(p_2) | \gamma^*(q_1,\lambda_1) \gamma^*(q_2,\lambda_2) \> \nn
		&= - e^2 \epsilon_\mu^{\lambda_1}(q_1) \epsilon_\nu^{\lambda_2}(q_2) \int d^4x \, d^4y \, e^{-i (q_1 \cdot x + q_2 \cdot y)} \<\pi^a(p_1) \pi^b(p_2)| T \{ j_\mathrm{em}^\mu(x) j_\mathrm{em}^\nu(y) \} | 0 \> \nn
		&= - e^2 (2\pi)^4 \delta^{(4)}(p_1+p_2-q_1-q_2) \epsilon_\mu^{\lambda_1}(q_1) \epsilon_\nu^{\lambda_2}(q_2) \nn
			&\quad \times \int d^4x \, e^{-i q_1 \cdot x} \<\pi^a(p_1) \pi^b(p_2)| T \{ j_\mathrm{em}^\mu(x) j_\mathrm{em}^\nu(0) \} | 0 \>  \nn
		&= i e^2 (2\pi)^4 \delta^{(4)}(p_1+p_2-q_1-q_2) \epsilon_\mu^{\lambda_1}(q_1) \epsilon_\nu^{\lambda_2}(q_2) W^{\mu\nu}_{ab}(p_1,p_2,q_1)
\end{align}
and the contraction with polarization vectors finally defines the helicity amplitudes according to
\begin{align}
	 \epsilon_\mu^{\lambda_1}(q_1) \epsilon_\nu^{\lambda_2}(q_2) W^{\mu\nu}_{ab}(p_1,p_2,q_1) = e^{i(\lambda_1-\lambda_2)\phi} H_{\lambda_1\lambda_2}^{ab}.
\end{align}

Next, the kinematic invariants are taken as\footnote{We denote $\gamma^*\gamma^*\to\pi\pi$ as the $s$-channel process, which is the canonical choice in the context of HLbL. 
Note that in the literature on RS equations~\cite{Hite:1973pm,Buettiker:2003pp,Hoferichter:2011wk,Ditsche:2012fv,Hoferichter:2015hva} usually the elastic channel, here pion Compton scattering, is considered the $s$-channel.}
\begin{align}
	\begin{split}
		s &= (q_1+q_2)^2 = (p_1+p_2)^2, \\
		t &= (q_1-p_1)^2 = (q_2-p_2)^2, \\
		u &= (q_1-p_2)^2 = (q_2-p_1)^2,
	\end{split}
\end{align}
satisfying 
\begin{equation}
\label{onshell}
s+t+u = q_1^2 + q_2^2 + 2 M_\pi^2=\Sigma_{\pi\pi}.
\end{equation} 
For the helicity amplitudes it is also convenient to choose a frame, 
we construct the helicity amplitudes with the momenta and polarization vectors in the $s$-channel center-of-mass system. This gives
\begin{align}
	\begin{split}
		q_1 &= ( E_{q_1}, 0, 0, |\vec q| ) , \qquad q_2 = ( E_{q_2}, 0, 0, -|\vec q| ) , \\
		p_1 &= ( E_p, |\vec p| \sin \theta \cos\phi, |\vec p| \sin \theta \sin\phi, |\vec p| \cos\theta ) , \\
		p_2 &= ( E_p, -|\vec p| \sin \theta \cos\phi, -|\vec p| \sin \theta \sin\phi, -|\vec p| \cos\theta ) ,
	\end{split}
\end{align}
where
\begin{align}
	\begin{split}
		E_{q_1} &= \sqrt{ q_1^2 + \vec q^2} = \frac{s + q_1^2 - q_2^2}{2\sqrt{s}}, \qquad E_{q_2} = \sqrt{ q_2^2 + \vec q^2} = \frac{s - q_1^2 + q_2^2}{2\sqrt{s}}, \qquad |\vec q| = \frac{\lambda^{1/2}(s,q_1^2,q_2^2)}{2\sqrt{s}} , \\
		E_p &= \sqrt{ M_\pi^2 + \vec p^2} = \frac{\sqrt{s}}{2} , \qquad |\vec p| = \sqrt{\frac{s}{4} - M_\pi^2} = \frac{\sqrt{s}}{2} \sigma_\pi(s) .
	\end{split}
\end{align}
and we introduced the notation
\begin{align}
	\sigma_\pi(s) &= \sqrt{ 1 - \frac{4M_\pi^2}{s}}, \qquad \lambda(a,b,c) = a^2 + b^2 + c^2 - 2(a b + b c + c a ),
\end{align}
the $s$-channel scattering angle
\begin{align}
	\label{eq:CosTheta}
	z = \cos\theta = \frac{t-u}{4 |\vec q| |\vec p|} = \frac{t-u}{\sigma_\pi(s) \lambda^{1/2}(s,q_1^2,q_2^2)},
\end{align}
as well as the polarization vectors
\begin{align}
	\begin{split}
		\epsilon_\pm(q_1) &= \mp \frac{1}{\sqrt{2}}( 0, 1, \pm i, 0 ) , \\
		\epsilon_0(q_1) &= \frac{1}{\xi_1}( |\vec q|, 0, 0, E_{q_1} ) , \\
		\epsilon_\pm(q_2) &= \mp \frac{1}{\sqrt{2}}( 0, 1, \mp i, 0 ) , \\
		\epsilon_0(q_2) &= \frac{1}{\xi_2}( -|\vec q|, 0, 0, E_{q_2} ) .
	\end{split}
\end{align}
For the particular choice of $\xi_i = \sqrt{q_i^2}$ the longitudinal states are normalized to $1$, but, since the off-shell photons are not physical states, the choice of $\xi_i$ cannot affect
physical observables, a useful check on the calculation. For convenience, we also define helicity amplitudes that stay finite in the limit $q_i^2 \to 0$
\begin{align}
\label{def_kappa}
	H_{\lambda_1\lambda_2} = \kappa_{\lambda_1}^1 \kappa_{\lambda_2}^2 \bar H_{\lambda_1\lambda_2}, \qquad \kappa_\pm^i = 1, \qquad \kappa_0^i = \frac{q_i^2}{\xi_i},
\end{align}
and introduce the labeling
\begin{align}
	\bar H_{1} = \bar H_{++}, \qquad \bar H_{2} = \bar H_{+-}, \qquad \bar H_{3} = \bar H_{+0} + \bar H_{0+}, \qquad \bar H_{4} = \bar H_{+0} - \bar H_{0+}, \qquad \bar H_{5} = \bar H_{00}.
\end{align}
Finally, the helicity amplitudes are expanded into partial waves according to~\cite{Jacob:1959at}
\begin{align}
	\bar H_{\lambda_1\lambda_2} = \sum_J (2J+1) d^J_{m0}(z) h_{J,\lambda_1\lambda_2}(s),
\end{align}
where $m = |\lambda_1 - \lambda_2|$.

\subsection{Tensor decomposition and dispersion relations}
\label{sec:dispersion_relations}

DRs should not be derived for the helicity amplitudes directly due to their complicated analytic structure, but instead for scalar functions that are free of kinematic
singularities and zeros. Such a basis has been derived in~\cite{Colangelo:2015ama} following the general recipe established in~\cite{Bardeen:1969aw}. In contrast to the singly-virtual case, however,
the doubly-virtual process is sufficiently complicated that an additional limitation first observed in nucleon Compton scattering~\cite{Tarrach:1975tu} occurs, i.e.\ that to cover
all kinematic limits a sixth Lorentz structure needs to be provided, in addition to the five expected in correspondence to the five independent helicity amplitudes.
Fortunately, the number of required scalar functions can be reduced by using crossing symmetry in the pion system, again in analogy to nucleon Compton scattering~\cite{Drechsel:1997xv}, finally
leading to the scalar functions $A_i$ as defined in~\cite{Colangelo:2015ama}. Explicitly, we have
\begin{equation}
 W_{\mu\nu} = \sum_{i=1}^5 T^i_{\mu\nu} A_i,
\end{equation}
with
\begin{align}
	\begin{split}
		T_1^{\mu\nu} &= q_1 \cdot q_2 g^{\mu\nu} - q_2^\mu q_1^\nu , \\
		T_2^{\mu\nu} &= q_1^2 q_2^2 g^{\mu\nu} + q_1 \cdot q_2 q_1^\mu q_2^\nu - q_1^2 q_2^\mu q_2^\nu - q_2^2 q_1^\mu q_1^\nu , \\
		T_3^{\mu\nu} &= (t-u)(\tilde T_3^{\mu\nu} - \tilde T_4^{\mu\nu}), \\
		T_4^{\mu\nu} &= q_1 \cdot q_2 q_3^\mu q_3^\nu - \frac{1}{4} (t-u)^2  g^{\mu\nu} + \frac{1}{2} (t-u) \left( q_3^\mu q_1^\nu - q_2^\mu q_3^\nu \right) , \\
		T_5^{\mu\nu} &= q_1^2 q_2^2 q_3^\mu q_3^\nu + \frac{1}{2}(t-u) \left(q_1^2 q_3^\mu q_2^\nu - q_2^2 q_1^\mu q_3^\nu \right) - \frac{1}{4} (t-u)^2 q_1^\mu q_2^\nu,
	\end{split}
\end{align}
where $q_3=p_2-p_1$ and
\begin{align}
	\begin{split}
		\tilde T_3^{\mu\nu} &= q_1 \cdot q_2 q_1^\mu q_3^\nu - q_1^2 q_2^\mu q_3^\nu - \frac{1}{2} (t-u) q_1^2 g^{\mu\nu} + \frac{1}{2} (t-u) q_1^\mu q_1^\nu , \\
		\tilde T_4^{\mu\nu} &= q_1 \cdot q_2 q_3^\mu q_2^\nu - q_2^2 q_3^\mu q_1^\nu + \frac{1}{2} (t-u) q_2^2 g^{\mu\nu} - \frac{1}{2} (t-u) q_2^\mu q_2^\nu.
	\end{split}
\end{align}
In terms of these functions, the helicity amplitudes read as follows
\begin{align*}
	\label{eq:ggpipiHelicityAmplitudes}
	\bar H_{++}&= -\frac{1}{2}(s-q_1^2-q_2^2) A_1 - q_1^2 q_2^2 A_2 + \frac{1}{2s}(s-4M_\pi^2)\lambda_{12}(s)z^2( q_1^2 + q_2^2 ) A_3 \\
		& \quad + \frac{1}{4} (s-4M_\pi^2)\left( (s - q_1^2 - q_2^2) + \left( \frac{(q_1^2-q_2^2)^2}{s} - (q_1^2 + q_2^2)\right) z^2 \right) A_4 \\
		& \quad + \frac{1}{2} q_1^2 q_2^2 (s-4M_\pi^2) (1-z^2) A_5 , \\
	\bar H_{+-} &= - \frac{1}{4} (s-4M_\pi^2) (1-z^2) \bigg( (s-q_1^2-q_2^2) A_4 + 2 q_1^2 q_2^2 A_5 \bigg) , \\
	\bar H_{+0} &= \frac{1}{4} \sqrt{\frac{2}{s}} (s-4M_\pi^2) z \sqrt{1-z^2} \bigg( \lambda_{12}(s) A_3 - (s+q_1^2-q_2^2) A_4  - q_1^2 (s-q_1^2 + q_2^2) A_5 \bigg) , \\
	\bar H_{0+}&= \frac{1}{4} \sqrt{\frac{2}{s}} (s-4M_\pi^2) z \sqrt{1-z^2} \bigg( \lambda_{12}(s) A_3  - (s-q_1^2+q_2^2) A_4  - q_2^2 (s + q_1^2 - q_2^2) A_5 \bigg) , \\
	\bar H_{00} &= \begin{aligned}[t]
		& - A_1 - \frac{1}{2}(s - q_1^2 - q_2^2) A_2 - \frac{1}{s}(s-4M_\pi^2)\lambda_{12}(s)z^2 A_3 \\
		& + (s-4M_\pi^2) z^2 A_4 + \frac{1}{4s} (s - 4M_\pi^2) \left(s^2 - (q_1^2 - q_2^2)^2\right) z^2 A_5, \end{aligned} \mytag
\end{align*}
where $\lambda_{12}(s) = \lambda(s,q_1^2,q_2^2)$. In this paper, we will make repeated reference to the expressions that follow from the pion-pole terms as well as the tree-level exchange
of vector mesons, with partial waves $N_{J,i}(s)$ and $h_{J,i}^V(s)$, see App.~\ref{app:Bornterms}.

The form of the DRs for the coefficient functions $A_i$ is defined by a second constraint on the Mandelstam variables besides the on-shell condition~\eqref{onshell}.
As argued in~\cite{Hite:1973pm}, the optimal choice for a process with crossing properties of $\gamma^*\gamma^*\to\pi\pi$
is given by hyperbolic DRs (HDRs), for which the dispersive variables are constrained to lie on hyperbolas of the form
\begin{align}
\label{HDR_conditions}
	(t-a)(u-a) = (t'-a)(u'-a) = b, \qquad s + t + u = s' + t' + u' = q_1^2 + q_2^2 + 2 M_\pi^2,
\end{align}
which implies the relation
\begin{align}
	\frac{ds'}{s'-s} = dt' \left( \frac{1}{t^\prime-t} + \frac{1}{t^\prime-u} - \frac{1}{t^\prime-a} \right)
\end{align}
for the differentials
and
\begin{align}
	z = \frac{t-u}{\sigma_\pi(s) \lambda^{1/2}_{12}(s)}, \qquad z' = \frac{t'-u'}{\sigma_\pi(s') \lambda^{1/2}_{12}(s')},
\end{align}
for the scattering angles, leading to 
\begin{align}
\label{HDRs}
	\begin{split}
		A_i(s,t,u) &= A_i^{\pi}(s,t,u) + \frac{1}{\pi} \int_{4M_\pi^2}^\infty ds^\prime \frac{\Im A_i(s^\prime, z^\prime)}{s^\prime - s} \\
			&\quad + \frac{1}{\pi} \int_{t_0}^\infty dt^\prime \Im A_i(t^\prime,u^\prime) \left( \frac{1}{t^\prime-t} + \frac{1}{t^\prime-u} - \frac{1}{t^\prime-a} \right),
	\end{split}
\end{align}
with Born terms $A_i^{\pi}$ given in App.~\ref{app:Bornterms}.
In writing~\eqref{HDRs} we have implicitly assumed that the Born-subtracted amplitudes fulfill unsubtracted HDRs. 
We stress that this is not equivalent to assuming unsubtracted HDRs for the full amplitude, 
since for $A_1$ the Born term itself goes asymptotically to a constant along the hyperbola. 
Assuming unsubtracted HDRs for the full amplitude would thus imply cancellations of this constant behavior with a contribution from heavier intermediate states.

A realistic description of $\gamma^*\gamma^*\to\pi\pi$ beyond the $S$-waves requires further contributions to the LHC, most importantly the exchange of vector mesons,
see App.~\ref{app:Bornterms} for the explicit expressions. This Lagrangian-based representation suffers from a polynomial ambiguity~\cite{GarciaMartin:2010cw}: 
choosing a different Lagrangian representation alters the real part of the amplitude, while the residues of the vector-resonance poles are free from such ambiguities. 
In the narrow-width limit, in which the imaginary parts of the vector-meson exchange collapse to $\delta$-functions, this can be demonstrated by comparing the expression resulting from the HDRs~\eqref{HDRs}
with the starting point~\eqref{Ai_V}. We find the differences
\begin{align}
	\begin{split}
		\Delta A_1^{V,\mathrm{HDR}} &= C_V^2 F_{V\pi}(q_1^2) F_{V\pi}(q_2^2) \left( \frac{(t-M_V^2)(u-M_V^2)}{2(a-M_V^2)^2} + \frac{s-3M_\pi^2-M_V^2}{a-M_V^2} - 4 \right), \\
		\Delta A_2^{V,\mathrm{HDR}} &= - \Delta A_4^{V,\mathrm{HDR}}  = - C_V^2 F_{V\pi}(q_1^2) F_{V\pi}(q_2^2) \frac{1}{a-M_V^2}, \\
		\Delta A_3^{V,\mathrm{HDR}} &= \Delta A_5^{V,\mathrm{HDR}} = 0.
	\end{split}
\end{align}
where $\Delta A_i^{V,\mathrm{HDR}} = A_i^{V,\mathrm{HDR}} - A_i^V$. In the limit $a\to\infty$ most of these differences disappear
\begin{align}
	\begin{split}
		\lim_{a\to\infty} \Delta A_1^{V,\mathrm{HDR}} &= - 4 C_V^2 F_{V\pi}(q_1^2) F_{V\pi}(q_2^2), \\
		\lim_{a\to\infty} \Delta A_i^{V,\mathrm{HDR}} &= 0, \qquad i\in\{2,3,4,5\}.
	\end{split}
\end{align}
The remaining ambiguity maps onto the polynomial obtained when changing the representation of the vector mesons from vector to antisymmetric tensor fields~\cite{GarciaMartin:2010cw,Ecker:1988te,Ecker:1989yg},
and due to~\eqref{eq:ggpipiHelicityAmplitudes} only affects the $S$-waves. 
In the following, we will indeed define the resonance LHCs by their $a\to\infty$ limit,  which corresponds to a fixed-$s$ DR, because this 
is the situation encountered in a dispersive approach to HLbL scattering where the different topologies are defined using 
the Mandelstam representation. Moreover, we will not comment further on the $S$-wave case---there, the consideration of subtractions is unavoidable to capture model-independently the effect of
vector resonances in the LHC---but concentrate on how to extend the dispersive description to $D$-waves.

The basic idea in the derivation of RS equations is then as follows: expand the imaginary parts of~\eqref{HDRs} into partial waves, express the internal angle $z'$ in terms of the external angle $z$ by means of the hyperbola conditions~\eqref{HDR_conditions}, and project the whole system onto partial waves. In contrast to~\cite{Hoferichter:2011wk} we will not calculate 
the kernel functions for the LHC explicitly, but directly work with a narrow-width approximation for the resonances. The resulting system of partial-wave DRs then takes the form
\begin{align}
	\label{eq:PartialWavesBornResResc}
		h_{J,i}(s) = N_{J,i}(s) + h_{J,i}^{V,\text{fixed-}s}(s)
			+ \sum_{J'} \sum_j \frac{1}{\pi} \int_{4M_\pi^2}^\infty ds^\prime K_{JJ'}^{ij}(s,s') \Im h_{J',j}(s'),
\end{align}
with $s$-channel kernel functions $K_{JJ'}^{ij}(s,s')$. The calculation of these kernel functions is straightforward, but reveals ostensible singularities in $1/s$ as well as factors involving $\sqrt{s}$ that originate from the definition of the helicity amplitudes. Before turning to the explicit form of the kernel functions, we therefore first study singularities that may be produced by the partial-wave expansion.

\subsection{Kinematic singularities and partial-wave expansion}
\label{sec:singularities_partial_wave}

Using the recipe of~\cite{Martin:1970}, the kinematic singularities in the helicity amplitudes can be separated according to
\begin{align*}
	\tilde H_{1} &= \lambda_{12}(s) \bar H_{1}, \\
	\tilde H_{2} &= \frac{1}{s-4M_\pi^2} \frac{1}{1-z^2} \bar H_{2}, \\
	\tilde H_{3} &= \frac{\lambda_{12}^{1/2}(s)}{\sqrt{s-4M_\pi^2}} \frac{1}{\sqrt{1-z^2}} \bar H_{3}, \\
	\tilde H_{4} &= \frac{\lambda_{12}^{1/2}(s)}{\sqrt{s-4M_\pi^2}} \frac{1}{\sqrt{1-z^2}} \bar H_{4}, \\
	\tilde H_{5} &= \lambda_{12}(s) \bar H_{5},
	\mytag
\end{align*}
and indeed the $\tilde H_i$ are given by a sum of the $A_i$ with coefficient functions that are polynomials in $s$, $t$, $u$.
However, we are mainly interested in the kinematic singularities of the partial waves, not the full amplitudes. To derive the corresponding singularities---with critical points $s=0$, $s=4\mpi^2$, and the zeros of $\lambda_{12}(s)$---let us assume that the scalar functions $A_i$ fulfill an unsubtracted fixed-$s$ DR
\begin{align}
	A_i(s,t) = \frac{1}{\pi} \int dt' \frac{\Im A_i(s,t')}{t'-t}.
\end{align} 
By performing the angular integrals in terms of Legendre functions of the second kind, in analogy to the partial-wave projection in App.~\ref{app:Bornterms}, this leads to the representation
\begin{align*}
	\label{eq:HelicityPartialWavesUnsubtractedFixedS}
	h_{J,1}(s) &= \frac{1}{\pi} \int dt' \frac{2}{\sigma_\pi(s) \lambda_{12}^{1/2}(s)} \begin{aligned}[t] & \bigg[ Q_J(x_{t'}) \begin{aligned}[t]  & \bigg( - \frac{s-q_1^2-q_2^2}{2} \Im A_1(s,t') - q_1^2 q_2^2 \Im A_2(s,t') \\
																	&+ \frac{(s-4M_\pi^2)(s-q_1^2-q_2^2)}{4} \Im A_4(s,t') \\
																	&+ \frac{q_1^2 q_2^2 (s-4M_\pi^2)}{2} \Im A_5(s,t') \bigg) \end{aligned} \\
																	& + ( x_{t'}^2 Q_J(x_{t'}) - x_{t'} \delta_{J0} ) \sigma_\pi^2(s) \\
																		&\quad \times \begin{aligned}[t] & \bigg( 
																		\frac{(q_1^2 + q_2^2) \lambda_{12}(s)}{2} \Im A_3(s,t') \\
																		&+ \frac{(q_1^2 - q_2^2)^2 - s(q_1^2+q_2^2)}{4} \Im A_4(s,t') - \frac{s q_1^2 q_2^2}{2} \Im A_5(s,t')
																	\bigg) 
																	 \bigg] , \end{aligned} \end{aligned} \\
	h_{J,2}(s) &= \frac{1}{\pi} \int dt' \frac{2(s-4M_\pi^2)}{\sigma_\pi(s) \lambda_{12}^{1/2}(s)}  \sqrt{\frac{(J+2)!}{(J-2)!}}  \bigg[ \frac{Q_{J-2}(x_{t'})}{(2J-1)(2J+1)} - \frac{2 Q_{J}(x_{t'})}{(2J-1)(2J+3)} + \frac{Q_{J+2}(x_{t'})}{(2J+1)(2J+3)} \bigg] \\*
				&\qquad \times \bigg[ -\frac{s-q_1^2-q_2^2}{4} \Im A_4(s,t') - \frac{q_1^2 q_2^2}{2} \Im A_5(s,t') \bigg] , \\
	h_{J,3}(s) &= \frac{1}{\pi} \int dt' \frac{1}{\sqrt{2s}} \frac{2(s-4M_\pi^2)}{\sigma_\pi(s) \lambda_{12}^{1/2}(s)}  \sqrt{\frac{J}{J+1}} x_{t'} \Big[ x_{t'} Q_J(x_{t'}) - Q_{J-1}(x_{t'}) \Big] \\*
				&\qquad \times \bigg[ \lambda_{12}(s) \Im A_3(s,t') - s \Im A_4(s,t') + \frac{(q_1^2-q_2^2)^2-s(q_1^2+q_2^2)}{2} \Im A_5(s,t') \bigg], \\
	h_{J,4}(s) &= \frac{1}{\pi} \int dt' \frac{1}{\sqrt{2s}} \frac{2(s-4M_\pi^2)}{\sigma_\pi(s) \lambda_{12}^{1/2}(s)}  \sqrt{\frac{J}{J+1}} x_{t'} \Big[ x_{t'} Q_J(x_{t'}) - Q_{J-1}(x_{t'}) \Big] \\*
				&\qquad \times (q_1^2 - q_2^2) \bigg[ - \Im A_4(s,t') - \frac{s - q_1^2-q_2^2}{2} \Im A_5(s,t') \bigg] , \\
	h_{J,5}(s) &=  \frac{1}{\pi} \int dt' \frac{2}{\sigma_\pi(s) \lambda_{12}^{1/2}(s)} \begin{aligned}[t] & \bigg[ Q_J(x_{t'}) \begin{aligned}[t]  & \bigg( - \Im A_1(s,t') - \frac{s-q_1^2- q_2^2}{2} \Im A_2(s,t') \bigg) \end{aligned} \\
																	& + ( x_{t'}^2 Q_J(x_{t'}) - x_{t'} \delta_{J0} ) \sigma_\pi^2(s) \\
																		&\quad \times \begin{aligned}[t] & \bigg( 
																		-\lambda_{12}(s) \Im A_3(s,t') + s \Im A_4(s,t') \\
																		&+ \frac{(s-q_1^2+q_2^2)(s+q_1^2-q_2^2)}{4} \Im A_5(s,t')
																	\bigg) 
																	 \bigg], \end{aligned} \end{aligned}
	\mytag
\end{align*}
where
\begin{align}
	x_{t'} = \frac{s-\Sigma_{\pi\pi}+2t'}{\sigma_\pi(s) \lambda_{12}^{1/2}(s)}.
\end{align}
From these relations we may read off the kinematic singularities as follows: at threshold the combination
\begin{align}
	\tilde Q_J(s,t') = \frac{1}{\sigma_\pi(s) \lambda_{12}^{1/2}(s)} Q_J(x_{t'})
\end{align}
behaves as
\begin{align}
	\tilde Q_J(s,t') \sim \Big[ (s-4M_\pi^2) \lambda_{12}(s) \Big]^{J/2},
\end{align}
while at $s=0$, $s^{-1/2} \tilde Q_J(s,t')$ is finite. Since, in addition $x_{t'}/(\sigma_\pi(s) \lambda_{12}^{1/2}(s))\sim s$ for $s\to 0$, the $1/s$ singularities in $\sigma_\pi^2(s)$ cancel
and we find\footnote{The signs and factors in $h_{J,3}(s)$ and $h_{J,4}(s)$ have been chosen to simplify the form of the final kernel functions.}
\begin{align*}
	h_{J,1}(s) &= \frac{\Big[ (s-4M_\pi^2) \lambda_{12}(s) \Big]^{J/2}}{\lambda_{12}(s)^{(1-\delta_{J0})}} \, \tilde h_{J,1}(s), \\
	h_{J,2}(s) &= \frac{\Big[ (s-4M_\pi^2) \lambda_{12}(s) \Big]^{J/2}}{\lambda_{12}(s)} \, \tilde h_{J,2}(s), \\
	h_{J,3}(s) &= - \sqrt{\frac{s}{2}} \frac{\Big[ (s-4M_\pi^2) \lambda_{12}(s) \Big]^{J/2}}{\lambda_{12}(s)} \, \tilde h_{J,3}(s), \\
	h_{J,4}(s) &= - (q_1^2 - q_2^2) \sqrt{\frac{s}{2}} \frac{\Big[ (s-4M_\pi^2) \lambda_{12}(s) \Big]^{J/2}}{\lambda_{12}(s)} \, \tilde h_{J,4}(s), \\
	h_{J,5}(s) &= \frac{\Big[ (s-4M_\pi^2) \lambda_{12}(s) \Big]^{J/2}}{\lambda_{12}(s)^{(1-\delta_{J0})}} \, \tilde h_{J,5}(s),
	\mytag
\end{align*}
where the functions $\tilde h_{J,i}(s)$ are regular at the thresholds. $s^{-1/2} \tilde h_{J,i}(s)$, $i=1,2,5$, are finite at $s=0$ for $J\ge2$.
Further zeros are possible for specific contributions, but should be considered to be of dynamical origin~\cite{Jackson:1968rfn}.
The functions $\tilde h_{J,i}(s)$ have LHCs, encoded in the $Q_J(x_{t'})$, as well as the right-hand cuts from the direct-channel contribution in $A_i(s,t')$.
As a cross check on~\eqref{eq:HelicityPartialWavesUnsubtractedFixedS}, we recover the fixed-$s$ resonance partial waves $\lim\limits_{a\to\infty} h_{J,i}^{V,\mathrm{HDR}}(s)$
if the $\delta$-function imaginary parts of the narrow-width resonance amplitudes are inserted.

For the RS system~\eqref{eq:PartialWavesBornResResc}, the central conclusion of this derivation is that upon the rescaling
\begin{align*}
\label{tildehJi}
	h_{0,i}(s) &= \tilde h_{0,i}(s), \qquad i = 1, 5, \\
	h_{2,i}(s) &= (s-4M_\pi^2) \tilde h_{2,i}(s), \quad i = 1,2,5, \\
	h_{2,3}(s) &= - \sqrt{\frac{s}{2}} (s-4M_\pi^2) \tilde h_{2,3}(s), \\
	h_{2,4}(s) &= - (q_1^2 - q_2^2) \sqrt{\frac{s}{2}} (s-4M_\pi^2) \tilde h_{2,4}(s),
	\mytag
\end{align*}
of the $S$- and $D$-waves, the $\tilde h_{J,i}(s)$ do not have further kinematic singularities provided that the full amplitudes $A_i(s,t)$ satisfy unsubtracted fixed-$s$ DRs.
This situation changes once subtractions are introduced in the fixed-$s$ DR. In this case, both the subtraction polynomial and the dispersion integral display $1/s$ singularities, whose residues will cancel each other if sum rules exist that reinstate the unsubtracted version.
In the derivation of RS equations the amplitudes are expanded into partial waves both at the level of the hyperbolic dispersion integrals as well as in the integrands. 
This implies that the full amplitudes are approximated by a truncated partial-wave series, which spoils the asymptotic behavior in the crossed channel, 
so that unsubtracted fixed-$s$ DRs are no longer possible. Therefore, additional $1/s$ singularities may appear at any finite order in the partial-wave expansion, 
and these are precisely the singularities observed in the RS kernels in~\eqref{eq:PartialWavesBornResResc}.

\subsection{Kernel functions}

Motivated by the discussion in the preceding section, we write the RS kernels not for the $h_{J,i}$, but for the $\tilde h_{J,i}$ according to~\eqref{tildehJi}, replacing~\eqref{eq:PartialWavesBornResResc} by
\begin{align}
	\label{eq:tildePartialWavesBornResResc}
		\tilde h_{J,i}(s) = \tilde N_{J,i}(s) + \tilde h_{J,i}^{V,\text{fixed-}s}(s)
			+ \sum_{J'} \sum_j \frac{1}{\pi} \int_{4M_\pi^2}^\infty ds^\prime K_{JJ'}^{ij}(s,s') \Im \tilde h_{J',j}(s').
\end{align}
The $S$-wave kernel functions recover the corresponding results from~\cite{Colangelo:2015ama}
\begin{align*}
	K_{00}^{11}(s,s') &= K_{00}^{55}(s,s') = \frac{1}{s'-s} - \frac{s'-q_1^2-q_2^2}{\lambda_{12}(s')}, \\
	K_{00}^{15}(s,s') &= \frac{2 q_1^2 q_2^2}{\lambda_{12}(s')}, \qquad K_{00}^{51}(s,s') = \frac{2}{\lambda_{12}(s')},
	\mytag
\end{align*}
as do the diagonal $D$-wave kernels. The full list of non-vanishing kernel functions reads
\begin{align*}
\label{final_kernels}
	K_{22}^{22}(s,s') &= \frac{1}{s'-s} - \frac{s'-q_1^2-q_2^2}{\lambda_{12}(s')}, \qquad
	K_{22}^{44}(s,s') = \frac{s'}{s} \left( \frac{1}{s'-s} - \frac{s'-q_1^2-q_2^2}{\lambda_{12}(s')} \right), \\
	K_{22}^{24}(s,s') &= \frac{2 s' q_1^2 q_2^2}{\lambda_{12}(s')}, \qquad
	K_{22}^{42}(s,s') = \frac{2}{s \lambda_{12}(s')}, \\
	K_{22}^{32}(s,s') &= \frac{2}{\lambda_{12}(s')}, \qquad
	K_{22}^{33}(s,s') = \frac{s'}{s} \frac{\lambda_{12}(s)}{\lambda_{12}(s')} \frac{1}{s'-s}, \qquad
	K_{22}^{34}(s,s') = \frac{s'}{s} \frac{s(q_1^2+q_2^2)-(q_1^2-q_2^2)^2}{\lambda_{12}(s')}, \\
	K_{22}^{11}(s,s') &= K_{22}^{55}(s,s')  =  \frac{s'}{s} \frac{\lambda_{12}(s)}{\lambda_{12}(s')} \left( \frac{1}{s'-s} - \frac{s'-q_1^2-q_2^2}{\lambda_{12}(s')} \right), \\
	K_{22}^{15}(s,s') &= \frac{2 s' q_1^2 q_2^2}{s} \frac{\lambda_{12}(s)}{\lambda_{12}^2(s')}, \qquad
	K_{22}^{51}(s,s') = \frac{2s'}{s} \frac{\lambda_{12}(s)}{\lambda_{12}^2(s')}, \\
	K_{22}^{12}(s,s') &= \frac{\lambda_{12}(s') \Big( s(q_1^2+q_2^2)-(q_1^2-q_2^2)^2\Big) - \lambda_{12}(s) \Big( s'(q_1^2+q_2^2)-(q_1^2-q_2^2)^2\Big) }{\sqrt{6}s\lambda_{12}^2(s')}, \\
	K_{22}^{13}(s,s') &= \frac{s'}{s} \frac{\lambda_{12}(s)}{\lambda_{12}^2(s')} \frac{s'(q_1^2+q_2^2)-(q_1^2-q_2^2)^2}{\sqrt{6}}, \\
	K_{22}^{14}(s,s') &= \frac{s'}{s} \frac{2 s q_1^2 q_2^2 \lambda_{12}(s')  - (s'-q_1^2-q_2^2)(q_1^2-q_2^2)^2 \lambda_{12}(s) }{\sqrt{6}\lambda_{12}^2(s')}, \\
	K_{22}^{52}(s,s') &= \frac{2 s' \lambda_{12}(s) - 4 s \lambda_{12}(s') }{\sqrt{6}s\lambda_{12}^2(s')}, \qquad
	K_{22}^{53}(s,s') = -\sqrt{\frac{2}{3}} \frac{{s'}^2}{s} \frac{\lambda_{12}(s)}{\lambda_{12}^2(s')}, \\
	K_{22}^{54}(s,s') &= \frac{s'}{s} \frac{2 (q_1^2 - q_2^2)^2 \lambda_{12}(s)  - (s+q_1^2-q_2^2)(s-q_1^2+q_2^2) \lambda_{12}(s') }{\sqrt{6}\lambda_{12}^2(s')}.
	\mytag
\end{align*}
There is no coupling of $D$- to $S$-waves, i.e.\ $K_{20}^{ij}(s,s') = 0$, but the $S$-waves couple to the $D$-waves through the kernels $K_{02}^{ij}(s,s')$, which depend linearly on the hyperbola parameter $a$. 
In the basis of the $\tilde h_{J,i}$, these kernel functions are lengthy and therefore not reproduced here. We will give them in a more convenient basis in Sect.~\ref{sec:Diagonalization} and their role in the MO solution will be studied in detail in Sect.~\ref{sec:MO_solution_Dwaves}.

We remark that this pattern seems to persist at higher orders in the partial-wave expansion: we have calculated all the kernel functions for $J,J'\le8$ and found that $K_{JJ'}^{ij}(s,s') = 0$ for $J>J'$. 
The kernels with $J=J'$ do not depend on the hyperbola parameter $a$, while the $J<J'$ kernels are polynomials in $a$ of the order $(J'-J)/2$. 
This behavior was indeed obtained in~\cite{Ditsche:2012fv} and it should be possible to prove the same here with similar methods.

\subsection{Diagonalization of the kernel functions}
\label{sec:Diagonalization}

The DRs for the helicity partial waves that follow from the RS system are integral equations that relate the helicity partial waves with their imaginary parts. The equations have the form of an inhomogeneous Omnès problem~\cite{Omnes:1958hv,Muskhelishvili:1953} and we will discuss the solution in Sect.~\ref{sec:MO_solution}. The MO solution is most easily found by performing another change of basis that diagonalizes the system of equations for a given angular momentum $J$ and decouples the system into a set of independent equations in standard MO form.
The $S$- and $D$-wave basis changes are given by
\begin{align}
	\label{eq:Diagonalization}
	\check h_{0,1}(s) &= \frac{1}{s - s_-} \Big[ \tilde h_{0,1}(s) + \frac{s_+ - s_-}{4} \tilde h_{0,5}(s) \Big] , \nn
	\check h_{0,5}(s) &= \frac{1}{s - s_+} \Big[ \tilde h_{0,1}(s) - \frac{s_+ - s_-}{4} \tilde h_{0,5}(s) \Big] , \nn
	\check h_{2,1}(s) &= \frac{1}{(s - s_-)(s - s_+)} \bigg[ \tilde h_{2,2}(s)+\frac{s}{2}\Big(\frac{s_++s_-}{2}\tilde h_{2,4}(s)-\tilde h_{2,3}(s)\Big) \bigg], \nn
	\check h_{2,2}(s) &= \frac{1}{(s-s_-)^2(s-s_+)}
		\begin{aligned}[t]
			& \bigg[ s\Big(\tilde h_{2,1}(s)+\frac{s_+-s_-}{4}\tilde h_{2,5}(s)\Big)+\frac{s_-}{\sqrt{6}}\Big(\tilde h_{2,2}(s)-s\tilde h_{2,3}(s)\Big) \nn
			&+\frac{s}{4\sqrt{6}}\Big(s_-(s_-+3s_+)+s(s_+-s_-)\Big)\tilde h_{2,4}(s) \bigg] , \end{aligned} \nn
	\check h_{2,3}(s) &= \frac{1}{(s-s_+)^2(s-s_-)}
		\begin{aligned}[t]
			& \bigg[ s\Big(\tilde h_{2,1}(s)-\frac{s_+-s_-}{4}\tilde h_{2,5}(s)\Big)+\frac{s_+}{\sqrt{6}}\Big(\tilde h_{2,2}(s)-s\tilde h_{2,3}(s)\Big) \nn
		&+\frac{s}{4\sqrt{6}}\Big(s_+(s_++3s_-)-s(s_+-s_-)\Big)\tilde h_{2,4}(s) \bigg] , \end{aligned} \nn
	\check h_{2,4}(s) &= \frac{1}{s - s_-} \Big[ \tilde h_{2,2}(s)+\frac{s_+-s_-}{4}s\tilde h_{2,4}(s) \Big] , \nn
	\check h_{2,5}(s) &= \frac{1}{s - s_+} \Big[ \tilde h_{2,2}(s)-\frac{s_+-s_-}{4}s\tilde h_{2,4}(s) \Big] ,
\end{align}
where $s_\pm=\big(\sqrt{q_1^2}\pm\sqrt{q_2^2}\big)^2$. In the on-shell or singly-virtual case, the poles in the kinematic prefactors get canceled by the 
soft-photon zeros at $s=s_\pm=q^2$~\cite{Low:1958sn}.
By writing the basis change in matrix form
\begin{align}
	\check h_{J,i}(s) = A_J^{ij}(s) \tilde h_{J,j}(s) ,
\end{align}
we find that the kernels with $J=J'$ are diagonalized to Cauchy kernels:
\begin{align}
	A_J(s) K_{JJ}(s,s') A_J^{-1}(s') = \frac{1}{s'-s} , \qquad J = 0, 2 ,
\end{align}
hence the new functions fulfill DRs in standard MO form with inhomogeneities $\check\Delta_{J,i}$ that contain the LHCs and the couplings due to the off-diagonal kernels:
\begin{align}
	\label{eq:checkPartialWavesBornResResc}
		\check h_{J,i}(s) &= \check \Delta_{J,i}(s) + \frac{1}{\pi} \int_{4M_\pi^2}^\infty ds^\prime \frac{\Im \check h_{J,i}(s')}{s'-s} , \nn
		\check \Delta_{J,i}(s) &= N_{J,i}(s) + \check h_{J,i}^{V,\text{fixed-}s}(s)
			+ \sum_{J'>J} \sum_j \frac{1}{\pi} \int_{4M_\pi^2}^\infty ds^\prime \check K_{JJ'}^{ij}(s,s') \Im \check h_{J',j}(s') .
\end{align}
In the new basis, the kernels that couple the $S$- to the $D$-waves turn out to be very compact:
\begin{align}
	\label{eq:SDKernels}
	\check K_{02}^{11}(s,s') &= 2 s_- \check K_{02}^{15}(s,s') = -\frac{5 s_-}{\sqrt{6}} \Big[ \frac{4 \mpi^2 s_+}{s s'} - 1 \Big], \qquad
	\check K_{02}^{14}(s,s') = \check K_{02}^{55}(s,s') = \frac{5}{\sqrt{6}}, \nn
	\check K_{02}^{51}(s,s') &= 2 s_+ \check K_{02}^{54}(s,s') = -\frac{5 s_+}{\sqrt{6}} \Big[ \frac{4 \mpi^2 s_-}{s s'} - 1 \Big], \nn
	\check K_{02}^{12}(s,s') &= \check K_{02}^{53}(s,s') = 5 \Big[ - \frac{2\mpi^2 s_- s_+}{s s'} + 6a - 4 \mpi^2 - s_- - s_+ + s + s'  \Big],
\end{align}
where $a$ is the hyperbola parameter and the kernel functions not listed explicitly vanish.

As an alternative to~\eqref{eq:checkPartialWavesBornResResc}, the resonance LHC can be written in terms of a dispersion integral over its discontinuity, see Sect.~\ref{sec:resonance_partial_waves} for details. This results in a representation
\begin{align}
	\label{eq:checkPartialWavesBornResRescAlt}
		\check h_{J,i}(s) &= \check \Delta_{J,i}(s) + \frac{1}{\pi} \int_{-\infty}^0 ds^\prime \frac{\Im \check h_{J,i}^{V,\text{fixed-}s}(s')}{s'-s}  + \frac{1}{\pi} \int_{4M_\pi^2}^\infty ds^\prime \frac{\Im \check h_{J,i}(s')}{s'-s} , \nn
		\check \Delta_{J,i}(s) &= N_{J,i}(s)
			+ \sum_{J'>J} \sum_j \frac{1}{\pi} \int_{-\infty}^0 ds^\prime \check K_{JJ'}^{ij}(s,s') \Im \check h_{J',j}^{V,\text{fixed-}s}(s') \nn
			&\qquad\qquad + \sum_{J'>J} \sum_j \frac{1}{\pi} \int_{4M_\pi^2}^\infty ds^\prime \check K_{JJ'}^{ij}(s,s') \Im \check h_{J',j}(s'). 
\end{align}

\subsection{Asymptotic behavior and sum rules}
\label{sec:Asymptotics}

The DRs~\eqref{eq:checkPartialWavesBornResResc} are a direct consequence of the HDRs~\eqref{HDRs}, following upon partial-wave projection and the basis change~\eqref{eq:Diagonalization}. They can be written without any subtractions provided that the initial HDRs are unsubtracted. The pure rescattering contributions
\begin{align}
	\check h_{J,i}^\text{resc}(s) := \check h_{J,i}(s) - \check \Delta_{J,i}(s) &= \frac{1}{\pi} \int_{4M_\pi^2}^\infty ds^\prime \frac{\Im \check h_{J,i}(s')}{s'-s}
\end{align}
are functions that contain only the right-hand unitarity cut. If we make the additional assumption that not only $\check h_{J,i}^\text{resc}(s)$ but $s\, \check h_{J,i}^\text{resc}(s)$ vanishes for $s\to\infty$, then the DR
\begin{align}
	s \, \check h_{J,i}^\text{resc}(s) = \frac{1}{\pi} \int_{4M_\pi^2}^\infty ds' \frac{s'\, \Im \check h_{J,i}(s')}{s'-s}
\end{align}
holds, which in turn implies the sum rules
\begin{align}
	\label{eq:SumRules}
	\frac{1}{\pi} \int_{4M_\pi^2}^\infty ds' \Im \check h_{J,i}(s') = 0.
\end{align}
These sum rules are essential to justify unsubtracted Omnès representations in Sect.~\ref{sec:MO_solution}, however they need to be validated. 
Based on the general consideration of unsubtracted fixed-$s$ DRs for the scalar functions~\eqref{eq:HelicityPartialWavesUnsubtractedFixedS}, we expect an asymptotic behavior of the $D$-wave rescattering contribution of
\begin{align}
	\label{eq:DWaveAsymptotics}
	\check h_{2,i}^\text{resc}(s) \asymp \left\{ \begin{matrix}
									s^{-2} \log s , \qquad & i = 1, 4, 5 , \\
									s^{-3} \log s , \qquad & i = 2,3  ,
								\end{matrix} \right.
\end{align}
which implies two additional sum rules
\begin{align}
	\frac{1}{\pi} \int_{4M_\pi^2}^\infty ds' s' \, \Im \check h_{2,i}(s') = 0 , \qquad i = 2, 3 .
\end{align}
For the $S$-waves, we would expect a behavior $\asymp s^{-1}\log s$, which in general would require a subtraction in the Omn\`es representation, in line with the 
discussion in Sect.~\ref{sec:dispersion_relations}.
Due to these sum rules, most of the contributions from the off-diagonal kernels in~\eqref{eq:SDKernels} in fact vanish and only the simplified kernels
\begin{align}
	\label{eq:SDKernelsReduced}
	\check K_{02}^{11}(s,s') &= 2 s_- \check K_{02}^{15}(s,s') = \check K_{02}^{51}(s,s') = 2 s_+ \check K_{02}^{54}(s,s') \nn
		&= \frac{2}{\sqrt{6}} \check K_{02}^{12}(s,s') = \frac{2}{\sqrt{6}} \check K_{02}^{53}(s,s') = - \frac{20 \mpi^2 s_+ s_-}{\sqrt{6} s s'}
\end{align}
need to be taken into account. In particular, the dependence on the hyperbola parameter drops out, as has to happen to avoid an unphysical dependence on $a$. 
In cases where 
\begin{align}
	\check h_{J,i}^\text{resc}(0) = \frac{1}{\pi} \int_{4M_\pi^2}^\infty ds^\prime \frac{\Im \check h_{J,i}(s')}{s'}
\end{align}
vanishes, no couplings of $S$- to $D$-waves would survive at all.

As a special case we may consider the LHC resonance partial waves, for which the sum rules
\begin{align}
	 \frac{1}{\pi} \int_{-\infty}^0 ds' \frac{\Im \check h_{2,i}^{V,\text{fixed-}s}(s')}{s'} &= 0 , \qquad i = 1, 2, 3, 4, 5, \nn
	 \frac{1}{\pi} \int_{-\infty}^0 ds' \Im \check h_{2,i}^{V,\text{fixed-}s}(s') &= 0 , \qquad i = 1, 2, 3 , \nn
	 \frac{1}{\pi} \int_{-\infty}^0 ds' s' \, \Im \check h_{2,i}^{V,\text{fixed-}s}(s') &= 0 , \qquad i = 2, 3,
\end{align}
are indeed fulfilled, but the two sum rules
\begin{align}
	 \frac{1}{\pi} \int_{-\infty}^0 ds' \Im \check h_{2,i}^{V,\text{fixed-}s}(s') &\neq  0 , \qquad i = 4, 5 ,
\end{align}
are violated
since the asymptotic behavior of the resonance contribution is worse than what we assume for the rescattering contribution:
\begin{align}
	\label{eq:ResonancePWAsymptotics}
	\check h_{J,i}^{V,\text{fixed-}s}(s) \asymp \left\{ \begin{matrix}
									s^{-1} \log s , \qquad & J = 0, \; i = 1,5  , \\
									s^{-2} , \qquad & J = 2, \; i = 1 , \\
									s^{-1} , \qquad & J = 2, \; i = 4, 5 , \\
									s^{-3} \log s , \qquad & J = 2, \; i = 2,3.
								\end{matrix} \right.
\end{align}
However, the asymptotic behavior in~\eqref{eq:ResonancePWAsymptotics} still implies that all $S$- and $D$-waves fulfill unsubtracted DRs
\begin{align}
	\check h_{J,i}^{V,\text{fixed-}s}(s) = \frac{1}{\pi} \int_{-\infty}^0 ds^\prime \frac{\Im \check h_{J,i}^{V,\text{fixed-}s}(s')}{s'-s}.
\end{align}

	% !TEX root = ../DV_Paper.tex

\section{Muskhelishvili--Omn\`es solution}
\label{sec:MO_solution}

\subsection[MO solution: $S$-waves]{MO solution: $\boldsymbol{S}$-waves}
\label{sec:MO_solution_Swaves}

Since the functions defined in~\eqref{eq:Diagonalization} fulfill Watson's theorem~\cite{Watson:1954uc}
\beq
	\Im \check h_{J,i}(s) = \sin\delta_J(s) e^{-i\delta_J(s)} \check h_{J,i}(s) \theta\big(s-4\mpi^2\big),
\eeq
with $\pi\pi$ phase shifts $\delta_J(s)$, the solution to the MO problem can be given immediately in terms of the Omn\`es functions
\beq
\label{Omnes_def}
	\Omega_J(s)=\exp\Bigg\{\frac{s}{\pi}\int_{4\mpi^2}^\infty ds'\frac{\delta_J(s')}{s'(s'-s)}\Bigg\} .
\eeq
We start the discussion by considering the restricted $S$-wave system~\cite{Colangelo:2017qdm,Colangelo:2017fiz}. The MO solution has the form
\begin{align}
	\check h_{0,i}(s) &= \check\Delta_{0,i}(s) + \frac{\Omega_0(s)}{\pi} \int_{4\mpi^2}^\infty ds'\frac{\check\Delta_{0,i}(s') \sin\delta_0(s')}{|\Omega_0(s')|(s'-s)} ,
\end{align}
provided that $( \check h_{0,i}(s) - \check\Delta_{0,i}(s) ) / \Omega_0(s)$ tends to zero for $s\to\infty$. For a phase shift reaching asymptotically $\delta_0(s) \asymp \pi$, the Omn\`es function behaves as $\Omega_0(s) \asymp s^{-1}$, i.e.\ the sum rules~\eqref{eq:SumRules} are employed to write the MO solution without subtractions.
Performing the basis change back to the original helicity amplitudes leads to
\begin{align}
\label{MO_solution_Swaves}
 h_{0,1}(s)&=\Delta_{0,1}(s)+\frac{\Omega_0(s)}{\pi}\int_{4\mpi^2}^\infty ds'\frac{\sin\delta_0(s')}{|\Omega_0(s')|}\bigg[\bigg(\frac{1}{s'-s}-\frac{s'-q_1^2-q_2^2}{\lambda_{12}(s')}\bigg)\Delta_{0,1}(s')+\frac{2q_1^2q_2^2}{\lambda_{12}(s')}\Delta_{0,5}(s')\bigg],\notag\\
 h_{0,5}(s)&=\Delta_{0,5}(s)+\frac{\Omega_0(s)}{\pi}\int_{4\mpi^2}^\infty ds'\frac{\sin\delta_0(s')}{|\Omega_0(s')|}\bigg[\bigg(\frac{1}{s'-s}-\frac{s'-q_1^2-q_2^2}{\lambda_{12}(s')}\bigg)\Delta_{0,5}(s')+\frac{2}{\lambda_{12}(s')}\Delta_{0,1}(s')\bigg].
\end{align}
In~\cite{Colangelo:2017qdm,Colangelo:2017fiz} this solution was evaluated using the pion Born terms as LHCs and a $\pi\pi$ phase shift that cuts off the $f_0(980)$ and thus the coupling to the $K\bar K$ channel.
Phenomenologically, the pion-pole LHC produces the polarizabilities~\cite{Colangelo:2017qdm,Colangelo:2017fiz}
\begin{align}
 (\alpha_1-\beta_1)^{\pi^\pm,\pi\text{-pole LHC}}&=(5.4\ldots 5.8)\times
 10^{-4}\,\text{fm}^3, \notag \\
 (\alpha_1-\beta_1)^{\pi^0,\pi\text{-pole LHC}}&=(11.2\ldots 8.9)\times
 10^{-4}\,\text{fm}^3,
\end{align}
for the charged pion in perfect agreement with the chiral $2$-loop prediction $5.7(1.0)\times
 10^{-4}\,\text{fm}^3$~\cite{Gasser:2006qa} as well as the 
COMPASS measurement $4.0(1.2)_\text{stat}(1.4)_\text{syst}\times 10^{-4}\,\text{fm}^3$~\cite{Adolph:2014kgj}.
For the neutral pion the chiral prediction $-1.9(2)\times 10^{-4}\,\text{fm}^3$~\cite{Gasser:2005ud} is much smaller, a discrepancy explained
by the fact that the neutral channel is much stronger affected by the contribution from vector-meson exchange~\cite{Tanabashi:2018oca} 
\beq
\frac{\Gamma_\omega\times \BR[\omega\to\pi^0\gamma]+\Gamma_\rho \times
  \BR[\rho^0\to\pi^0\gamma]}{\Gamma_\rho \times \BR[\rho^\pm\to\pi^\pm\gamma]}\sim
12. 
\eeq
Therefore, the discrepancy does not necessarily point at a violation of the sum rule~\eqref{eq:SumRules}, but rather the approximation of the LHC by the pion Born term only. 
However, as argued in Sect.~\ref{sec:dispersion_relations}, to get the phenomenology of the neutral-pion dipole polarizabilities right the introduction of subtractions is nevertheless unavoidable,
otherwise the vector-meson LHC remains ambiguous. For the quadrupole polarizabilities the vector-meson contribution indeed restores agreement with ChPT even for the neutral pion~\cite{Colangelo:2017fiz}.  

For the $D$-waves, polynomial ambiguities in the vector-meson LHCs do not occur, so that unsubtracted DRs in principle become possible. To this end, a modified MO solution was derived
in~\cite{GarciaMartin:2010cw} in which the vector mesons are not included via the inhomogeneities, but directly in terms of their partial waves. This corresponds to the MO solution of the DR~\eqref{eq:checkPartialWavesBornResRescAlt}, which in the $S$-wave case leads to a modification of~\eqref{MO_solution_Swaves} according to
\begin{align}
\label{MO_solution_Swaves_Bachir}
 h_{0,1}(s)&=N_{0,1}(s)
 +\frac{\Omega_0(s)}{\pi}\int_{-\infty}^{0} \frac{ds'}{\Omega_0(s')}\bigg[\bigg(\frac{1}{s'-s}-\frac{s'-q_1^2-q_2^2}{\lambda_{12}(s')}\bigg)\Im h^V_{0,1}(s')+\frac{2q_1^2q_2^2}{\lambda_{12}(s')}\Im h^V_{0,5}(s')\bigg]\notag\\
 &+\frac{\Omega_0(s)}{\pi}\int_{4\mpi^2}^\infty ds'\frac{\sin\delta_0(s')}{|\Omega_0(s')|}\bigg[\bigg(\frac{1}{s'-s}-\frac{s'-q_1^2-q_2^2}{\lambda_{12}(s')}\bigg)N_{0,1}(s')+\frac{2q_1^2q_2^2}{\lambda_{12}(s')}N_{0,5}(s')\bigg],\notag\\
 h_{0,5}(s)&=N_{0,5}(s)
 +\frac{\Omega_0(s)}{\pi}\int_{-\infty}^{0}\frac{ds'}{\Omega_0(s')}\bigg[\bigg(\frac{1}{s'-s}-\frac{s'-q_1^2-q_2^2}{\lambda_{12}(s')}\bigg)\Im h^V_{0,5}(s')+\frac{2}{\lambda_{12}(s')}\Im h^V_{0,1}(s')\bigg]\notag\\
 &+\frac{\Omega_0(s)}{\pi}\int_{4\mpi^2}^\infty ds'\frac{\sin\delta_0(s')}{|\Omega_0(s')|}\bigg[\bigg(\frac{1}{s'-s}-\frac{s'-q_1^2-q_2^2}{\lambda_{12}(s')}\bigg)N_{0,5}(s')+\frac{2}{\lambda_{12}(s')}N_{0,1}(s')\bigg],
\end{align}
where the new integrals extend over the LHC, see Sect.~\ref{sec:resonance_partial_waves} for details. Although a subtracted DR was used for the numerical analysis in~\cite{GarciaMartin:2010cw},
it was shown that sum rules that would establish an unsubtracted version are nearly fulfilled, indicating that an approximate description should be possible based on an unsubtracted system as well.

\subsection[MO solution: $D$-waves]{MO solution: $\boldsymbol{D}$-waves}
\label{sec:MO_solution_Dwaves}

With the diagonalization of the $D$-wave system derived in Sect.~\ref{sec:Diagonalization}, the MO solutions follow immediately, as the defining DRs are given in decoupled form~\eqref{eq:checkPartialWavesBornResResc} or~\eqref{eq:checkPartialWavesBornResRescAlt}. The solution reads\footnote{For brevity, we will only quote the standard MO solutions in the following, with straightforward extensions to vector-resonance LHCs as in~\eqref{MO_solution_Swaves_Bachir}.}
\begin{align}
	\label{eq:MOSolutionDwaves}
	\check h_{2,i}(s) &= \check\Delta_{2,i}(s) + \frac{\Omega_2(s)}{\pi} \int_{4\mpi^2}^\infty ds'\frac{\check\Delta_{2,i}(s') \sin\delta_2(s')}{|\Omega_2(s')|(s'-s)} ,
\end{align}
where $\delta_2$ is the $D$-wave $\pi\pi$-scattering phase shift and $\Omega_2$ the corresponding Omn\`es function. From the functions $\check h_{2,i}$, we obtain the original helicity partial waves by inverting the basis change~\eqref{eq:Diagonalization}:
\begin{align}
	\label{eq:InverseBasisChangeDwaves}
	s\tilde h_{2,1}(s) &= - (s-s_+)(s-s_-) \bigg[ \frac{s_++s_-}{\sqrt{6}} \check h_{2,1}(s)
					- \frac{1}{2} \check h_{23}^+(s) \bigg] \nn
					&\quad + \frac{1}{4\sqrt{6}} \bigg[ (s_++s_-) \check h_{45}^+(s) + (s_+-s_-)^2 \check h_{45}^-(s) \bigg] , \nn
	\tilde h_{2,2}(s) &= \frac{1}{2} \check h_{45}^+(s) , \nn
	s\tilde h_{2,3}(s) &= -2(s-s_+)(s-s_-) \check h_{2,1}(s) + \check h_{45}^+(s) + (s_++s_-) \check h_{45}^-(s),\notag\\
	s\tilde h_{2,4}(s) &= 2 \check h_{45}^-(s),\notag\\
	s\tilde h_{2,5}(s) &= (s-s_+)(s-s_-) \bigg[ \frac{4}{\sqrt{6}} \check h_{2,1}(s) + 2 \check h_{23}^-(s) \bigg] - \frac{1}{\sqrt{6}} \bigg[ \check h_{45}^+(s) + (2s+s_++s_-) \check h_{45}^-(s) \bigg] ,
\end{align}
where we have introduced the combinations
\begin{align}
	\label{h_singly_offshell}
	\check h_{23}^+(s) &:= (s-s_-) \check h_{2,2}(s) + (s-s_+) \check h_{2,3}(s) , \quad  \check h_{45}^+(s) := (s-s_-) \check h_{2,4}(s) + (s-s_+) \check h_{2,5}(s) , \nn
	\check h_{23}^-(s) &:= \frac{(s-s_-) \check h_{2,2}(s) - (s-s_+) \check h_{2,3}(s)}{s_+-s_-} , \quad  \check h_{45}^-(s) := \frac{ (s-s_-) \check h_{2,4}(s) - (s-s_+) \check h_{2,5}(s)}{s_+-s_-} .
\end{align}
In the singly-virtual limit one has $s_+=s_-=q^2$, hence $\check h_{2,2}(s)= \check h_{2,3}(s)$, $\check h_{2,4}(s)=\check h_{2,5}(s)$, so that $\check h_{23}^-(s)$ and $\check h_{45}^-(s)$ remain finite. Their MO solution reads
\begin{align}
\label{MO_solution_23_45}
\check h_{23}^+(s) &= \check\Delta_{23}^+(s) + \frac{\Omega_2(s)}{\pi}\int_{4\mpi^2}^\infty ds' \frac{\sin\delta_2(s')}{|\Omega_2(s')|}\frac{1}{2}\bigg[\frac{(s_+-s_-)^2\check\Delta_{23}^-(s')}{\lambda_{12}(s')}
				+\bigg(\frac{s-s_+}{s'-s_+}+\frac{s-s_-}{s'-s_-}\bigg)\frac{\check\Delta_{23}^+(s')}{s'-s}\bigg] ,\nn
	\check h_{23}^-(s) &= \check\Delta_{23}^-(s) + \frac{\Omega_2(s)}{\pi}\int_{4\mpi^2}^\infty ds' \frac{\sin\delta_2(s')}{|\Omega_2(s')|}\frac{1}{2}\bigg[\frac{\check\Delta_{23}^+(s')}{\lambda_{12}(s')}
				+\bigg(\frac{s-s_+}{s'-s_+}+\frac{s-s_-}{s'-s_-}\bigg)\frac{\check\Delta_{23}^-(s')}{s'-s}\bigg] , \nn
\check h_{45}^+(s) &= \check\Delta_{45}^+(s) + \frac{\Omega_2(s)}{\pi}\int_{4\mpi^2}^\infty ds'\frac{\sin\delta_2(s')}{|\Omega_2(s')|}\frac{1}{2}\bigg[\frac{(s_+-s_-)^2\check\Delta_{45}^-(s')}{\lambda_{12}(s')}
				+\bigg(\frac{s-s_+}{s'-s_+}+\frac{s-s_-}{s'-s_-}\bigg)\frac{\check\Delta_{45}^+(s')}{s'-s}\bigg],\nn
	\check h_{45}^-(s) &= \check\Delta_{45}^-(s) + \frac{\Omega_2(s)}{\pi}\int_{4\mpi^2}^\infty ds'\frac{\sin\delta_2(s')}{|\Omega_2(s')|}\frac{1}{2}\bigg[\frac{\check\Delta_{45}^+(s')}{\lambda_{12}(s')}
				+\bigg(\frac{s-s_+}{s'-s_+}+\frac{s-s_-}{s'-s_-}\bigg)\frac{\check\Delta_{45}^-(s')}{s'-s}\bigg] ,
\end{align}
where $\check\Delta_{23}^\pm$ and $\check\Delta_{45}^\pm$ are defined in analogy to~\eqref{h_singly_offshell}. We also remark that for space-like virtualities, the zeros $s_\pm$ of $\lambda_{12}(s)$ need to be analytically continued according to
$s_\pm=-(\sqrt{-q_1^2}\pm \sqrt{-q_2^2})^2$.
The complete set of MO $D$-wave solutions is given by~\eqref{eq:MOSolutionDwaves}, together with the basis change~\eqref{eq:Diagonalization} and its inverse~\eqref{eq:InverseBasisChangeDwaves}. 
In particular, the solution~\eqref{MO_solution_23_45} amounts to a rewriting of the $S$-wave solution~\eqref{MO_solution_Swaves}, which can indeed be cast into the form~\eqref{MO_solution_23_45} once expressed in terms of
\begin{align}
\label{Swaves_inversion}
	\tilde h_{0,1}(s) &= \frac{1}{2} \check h_{15}^+(s) , \qquad \tilde h_{0,5}(s) = 2 \check h_{15}^-(s),\notag\\
	\check h_{15}^+(s) &:= (s-s_-) \check h_{0,1}(s) + (s-s_+) \check h_{0,5}(s) , \qquad
	\check h_{15}^-(s) := \frac{(s-s_-) \check h_{0,1}(s) - (s-s_+) \check h_{0,5}(s)}{s_+-s_-}.
\end{align}

 Finally, we turn to the $D$-wave contribution to the $S$-waves. According to~\eqref{eq:SDKernels}, these kernels produce an additional term in the $S$-wave inhomogeneities of the form
 \beq
  \check \Delta_{0,i}(s)=\frac{\check\alpha_{0,i}}{s},
 \eeq
 leading to
 \beq
 \check h_{0,i}(s)\to \check h_{0,i}(s)+\frac{\check\alpha_{0,i}}{s}\Omega_0(s),
 \eeq
 with
 \begin{align}
  \check\alpha_{0,1}&=-\frac{10\mpi^2s_+}{\sqrt{6}}\frac{1}{\pi}\int_{4\mpi^2}^\infty \frac{ds'}{s'}\Big(2s_-\Im \check h_{2,1}(s')+\sqrt{6} s_- \Im \check h_{2,2}(s')+\Im \check h_{2,5}(s')\Big),\notag\\
  \check\alpha_{0,5}&=-\frac{10\mpi^2s_-}{\sqrt{6}}\frac{1}{\pi}\int_{4\mpi^2}^\infty \frac{ds'}{s'}\Big(2s_+\Im \check h_{2,1}(s')+\sqrt{6} s_+ \Im \check h_{2,3}(s')+\Im \check h_{2,4}(s')\Big).
 \end{align}
For the inversion~\eqref{Swaves_inversion} one needs
\beq
\check h_{15}^\pm(s) \to \check h_{15}^\pm(s)+\frac{\check\alpha_{15}^\pm(s)}{s}\Omega_0(s),
\eeq
with
 \begin{align}
\check \alpha_{15}^+(s)&=-\frac{5\mpi^2}{\sqrt{6}}\frac{1}{\pi}\int_{4\mpi^2}^\infty \frac{ds'}{s'}
\bigg[4s_+s_-(2s-s_+-s_-)\Im \check h_{2,1}(s')\notag\\
&+\sqrt{6}s_+s_-\bigg(\bigg(\frac{s-s_+}{s'-s_+}+\frac{s-s_-}{s'-s_-}\bigg)\Im\check h_{23}^+(s')+\frac{(s_+-s_-)^2(s'-s)}{(s'-s_+)(s'-s_-)}\Im\check h_{23}^-(s')\bigg)\notag\\
&+\bigg(\frac{s_+(s-s_-)}{s'-s_+}+\frac{s_-(s-s_+)}{s'-s_-}\bigg)\Im\check h_{45}^+(s')+\frac{(s_+-s_-)^2(s_+s_--s's)}{(s'-s_+)(s'-s_-)}\Im\check h_{45}^-(s')\bigg],\notag\\
\check \alpha_{15}^-(s)&=-\frac{5\mpi^2}{\sqrt{6}}\frac{1}{\pi}\int_{4\mpi^2}^\infty \frac{ds'}{s'}
\bigg[4s_+s_-\Im \check h_{2,1}(s')\notag\\
&+\sqrt{6}s_+s_-\bigg(\frac{s'-s}{(s'-s_+)(s'-s_-)}\Im\check h_{23}^+(s')+\bigg(\frac{s-s_+}{s'-s_+}+\frac{s-s_-}{s'-s_-}\bigg)\Im\check h_{23}^-(s')\bigg)\notag\\
&-\frac{s_+s_--s's}{(s'-s_+)(s'-s_-)}\Im\check h_{45}^+(s')-\bigg(\frac{s_+(s-s_-)}{s'-s_+}+\frac{s_-(s-s_+)}{s'-s_-}\bigg)\Im\check h_{45}^-(s')\bigg].
 \end{align}
$\check \alpha_{15}^+(s)$ vanishes for $s_+=s_-=0$, so that the onshell process remains unaffected. 
As argued in Sect.~\ref{sec:singularities_partial_wave}, the appearance of the $1/s$ singularities is an artefact of the partial-wave expansion, and accordingly the size
of $\check \alpha_{15}^\pm(s)$ should be in line with effects expected from higher partial waves to allow for a cancellation in the full amplitude. 
We checked numerically that this residual coupling between $S$- and $D$-waves is indeed small, but this conclusion remains to be verified 
after integration over the weight functions in the $g-2$ integral. 
 
\subsection[Subtractions and the $f_2(1270)$ resonance]{\boldmath Subtractions and the $f_2(1270)$ resonance}
\label{sec:Subtractions}

In~\cite{GarciaMartin:2010cw} it was shown that sum rules for the subtraction constants in the modified Omn\`es representation are nearly fulfilled, making an approximate description possible that is based on an unsubtracted system. Surprisingly, the same observation does not hold for the $D$-wave analog of~\eqref{MO_solution_Swaves}, where the vector-meson LHC is treated as part of the inhomogeneity. Here, we want to clarify the reason why the two strategies lead to different results.

The assumption that an unsubtracted MO solution can be used relies in the standard Omn\`es representation on an asymptotic behavior
\begin{align}
	\frac{h^\text{std}(s) - \Delta(s)}{\Omega(s)} \asymp \frac{1}{s} ,
\end{align}
with $\Delta(s) = N(s) + h^V(s)$, whereas in the modified Omn\`es representation an unsubtracted DR is justified for
\begin{align}
	\frac{h^\text{mod}(s) - N(s)}{\Omega(s)} \asymp \frac{1}{s} .
\end{align}
If the Omn\`es function behaves as $\Omega(s) \asymp s^{-1}$, the two assumptions are equivalent provided that the vector-meson LHC vanishes asymptotically at least as $h^V(s) \asymp s^{-2}$. According to~\eqref{eq:ResonancePWAsymptotics}, this is not the case for the $S$-waves and the two $D$-waves $\check h^{V,\text{fixed-}s}_{2,i}$, $i=4,5$. The difference between the two representations is proportional to the Omn\`es function:
\begin{align}
	h^\text{std}(s) - h^\text{mod}(s) = \Omega(s) \bigg[ h^V(0) - \frac{1}{\pi} \int_{-\infty}^0 ds' \frac{\Im h^V(s')}{\Omega(s') s'} + \frac{1}{\pi} \int_{4\mpi^2}^\infty ds' \frac{h^V(s') \sin\delta(s')}{|\Omega(s')| s'} \bigg] .
\end{align}
In order for an unsubtracted standard MO solution to work, one would have to assume a cancellation of the bad high-energy behavior of the vector-meson LHC with the high-energy behavior of the rescattering contribution in $h^\text{std}(s)$, which seems unlikely. Therefore, the standard MO form is expected to work only when subtractions are introduced.
Note that the somewhat pathological high-energy behavior of the vector-meson LHC is only present in the real part, while the imaginary part is much better behaved. While the MO solution in the standard form keeps the bad high-energy behavior of the resonance LHC, the modified representation only involves the imaginary part of the vector-meson LHC and imposes a better high-energy behavior on the Born-subtracted partial waves. According to the general considerations of Sect.~\ref{sec:Asymptotics}, the asymptotic behavior for the $D$-wave rescattering~\eqref{eq:DWaveAsymptotics} should make an unsubtracted dispersion relation possible for the Born-subtracted part, hence a priori one would expect the modified MO solution to work even without subtractions.

Checking the MO solutions numerically, indeed it turns out that the unsubtracted standard MO form does not reproduce the peak of the narrow $f_2(1270)$ $D$-wave resonance. The effect on the resonance peak can be understood by noting that the modified MO solution is equivalent to the standard form with a subtraction, where the subtraction constant is 
effectively calculated in terms of the vector-meson LHC.
In this case, the resonance peak is fully described by the subtraction term: let us consider the part without subtraction constant,
\beq
\label{eq:SubtractionExample}
h(s)=\Delta(s) + \Omega(s) \frac{s}{\pi}\int_{4\mpi^2}^\infty ds'\frac{\Delta(s') \sin\delta(s')}{|\Omega(s')|s'(s'-s)},
\eeq
and let us further consider the simple case in which $\Delta(s)=\alpha/s$.
In this case, the dispersive integral can be performed analytically by using the spectral representation of the inverse Omn\`es function
\beq
\label{spectral_Omnes}
\Omega^{-1}(s)
=1-s \dot\Omega(0)-\frac{s^2}{\pi}\int_{4\mpi^2}^\infty ds'\frac{\sin\delta(s')}{|\Omega(s')|s'^2(s'-s)},
\eeq
yielding
\beq
\label{MO_standard_zero}
h(s)=\frac{\alpha}{s} \Omega(s)\big(1-s\dot\Omega(0)\big).
\eeq
The result is proportional to the Omn\`es function, as expected, but one finds an additional polynomial 
whose coefficients are determined by normalization and derivative of the Omn\`es function at $s=0$. For a narrow resonance with mass $M_R$, as the $f_2(1270)$ in the $D$-wave, 
one has $\Omega(s)\sim M_R^2/(M_R^2-s)$, so that $1-s\dot\Omega(0)$ vanishes at $s=M_R^2$. The resonance peak is thus described exclusively by the subtraction term that we dropped in~\eqref{eq:SubtractionExample}. Using an unsubtracted standard MO solution corresponds to fixing the subtraction constant with a sum rule that cannot be expected to hold and therefore leads to an incorrect description of the resonance.
Such a situation indeed occurs for some of the $D$-waves, which is why in the following we develop the formalism to include the vector mesons in the LHC in the modified MO representation as in~\eqref{MO_solution_Swaves_Bachir}
even in the doubly-virtual case, 
to be able to put forward an unsubtracted DR in the description of the $f_2(1270)$ resonance in $\gamma^*\gamma^*\to\pi\pi$. Avoiding the introduction of subtraction constants is 
advantageous for the generalization to the singly- or doubly-virtual case,
because otherwise their $q^2$-dependence would need to be addressed as well.

\subsection{Analytic structure of the resonance partial waves}
\label{sec:resonance_partial_waves}

To include the vector mesons directly in terms of the LHCs of their helicity partial waves, as in~\eqref{MO_solution_Swaves_Bachir}, 
we need to analyze the analytic structure of their LHCs in more detail. These LHCs are produced by the $t$- and $u$-channel exchange of a resonance with mass $M_V$.
For the cut structure itself the details of the partial-wave projection, i.e.\ angular momentum and helicity states, are irrelevant, let us therefore write symbolically
\begin{align}
	\label{eq:StructurePW}
	h^V(s) \sim \frac{1}{\sigma_\pi(s) \lambda_{12}^{1/2}(s)} Q(x_V),
\end{align}
where $Q(x)$ is a Legendre function of the second kind with a cut in the complex $x$-plane between $\pm1$ and $x_V$ is given in~\eqref{eq:HelAmpAbbreviations}.
Instead of considering the Legendre function, we can also study the angular integration path in the complex $t$- or $u$-plane, with endpoints at
\begin{align}
	t_\pm = u_\mp = \frac{1}{2} \left( q_1^2 + q_2^2 + 2M_\pi^2 - s \pm \sigma_\pi(s) \lambda_{12}^{1/2}(s) \right). 
\end{align}
In this way, the wrapping of the integration contour around the pole at $t,u=M_V^2$ determines possible singularities. Throughout, we will restrict the analysis
to space-like virtualities, $q_i^2 < 0$, as required for HLbL scattering. For time-like virtualities anomalous thresholds are certain to appear in any dispersive representation, even in MO solutions in the standard form~\eqref{MO_solution_Swaves}, see~\cite{Hoferichter:2013ama}.

First, possible kinematic square-root singularities at $s=4M_\pi^2$, $s=0$, and $s = - ( \sqrt{-q_1^2} \pm \sqrt{-q_2^2})^2 = s_\pm$ are in fact absent: in the explicit representation, these singularities of the Legendre function are 
balanced by the kinematic prefactor. Equivalently, in the path-deformation approach, they are lifted by the same factors coming from the Jacobian when switching from $z$ to $t$ or $u$ as integration variables. The only singularities are therefore logarithmic branch points at $x=\pm1$ or, equivalently $t_\pm = M_V^2$, given by
\begin{align}
	s_\mathrm{cut}^\pm = \frac{ M_\pi^2(2M_V^2+q_1^2+q_2^2) - M_\pi^4 -(M_V^2-q_1^2)(M_V^2-q_2^2) \pm \lambda^{1/2}(M_V^2,M_\pi^2,q_1^2) \lambda^{1/2}(M_V^2,M_\pi^2,q_2^2) }{2M_V^2}.
\end{align}
The other ends of the branch cuts are located at $s=0$ and $s=-\infty$, respectively, as can be inferred from the replacement $M_V^2\to\infty$. For $q_i^2\to0$, the two branch points are at
\begin{align}
	\lim_{q_i^2\to0} s_\mathrm{cut}^\pm = \left\{ \begin{matrix} 0 \\ - \frac{(M_V^2-M_\pi^2)^2}{M_V^2} \end{matrix} \right.,
\end{align}
hence only one branch cut from $-\infty$ to $- \frac{(M_V^2-M_\pi^2)^2}{M_V^2}$ is present, while the other one disappears. By writing the Källén function as
\begin{align}
	\lambda(M_V^2, M_\pi^2, q_i^2) = (q_i^2 - (M_V-M_\pi)^2) (q_i^2 - (M_V+M_\pi)^2),
\end{align}
it follows that the square roots in $s_\mathrm{cut}^\pm$ can only produce an imaginary part for 
\begin{align}
	q_i^2 \in \Big( (M_V-M_\pi)^2 , (M_V+M_\pi)^2 \Big),
\end{align}
i.e.\ for time-like virtualities. Therefore, for space-like virtualities the cut structure seems to remain simple: one expects just two branch cuts on the negative real axis, one from $-\infty$ to $s_\mathrm{cut}^-$, the other from $s_\mathrm{cut}^+$ to $0$.

However, an important subtlety arises that is reminiscent of anomalous thresholds in triangle diagrams, which appear for sufficiently large time-like virtualities: there, the discontinuity itself has singularities that depend on the virtualities and cross the unitarity cut of the triangle diagram. By entering the physical sheet, they require a deformation of the integration contour and add an ``anomalous'' discontinuity~\cite{Lucha:2006vc,Hoferichter:2013ama,Colangelo:2015ama}. Here, the discontinuity itself has the two square-root branch cuts from the kinematic factors in~\eqref{eq:StructurePW}, 
i.e.\ cuts for $s\in[0,4M_\pi^2]$ and $s\in[s_+,s_-]$. For space-like $q_i^2$, the second cut in the discontinuity lies between the two LHCs of the partial waves, 
i.e.\ $s_\pm \in[s_\mathrm{cut}^-,s_\mathrm{cut}^+]$, where the points $s_+$ and $s_\mathrm{cut}^-$ coincide for 
\beq
\label{virtualities_critical_point}
q_1^2q_2^2 = (M_V^2-M_\pi^2)^2.
\eeq
This condition can be fulfilled even for space-like virtualities, so that the corresponding points deserve special attention.

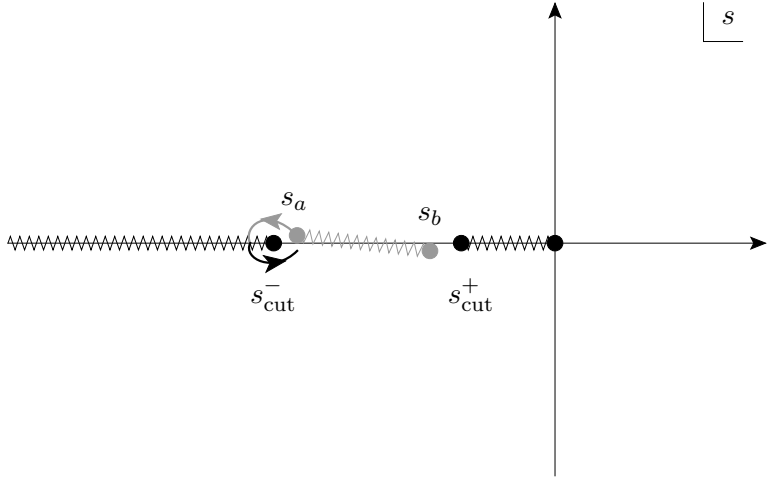
\begin{figure}[t]
	\centering
	\setlength{\unitlength}{1cm}
	\begin{picture}(10,6)(0,0)
		\put(9.4,6){$s$}
		\put(3.2,2.3){$s_\mathrm{cut}^-$}
		\put(5.8,2.3){$s_\mathrm{cut}^+$}
		\put(3.6,3.6){$s_a$}
		\put(5.4,3.4){$s_b$}
		\includegraphics[width=10cm]{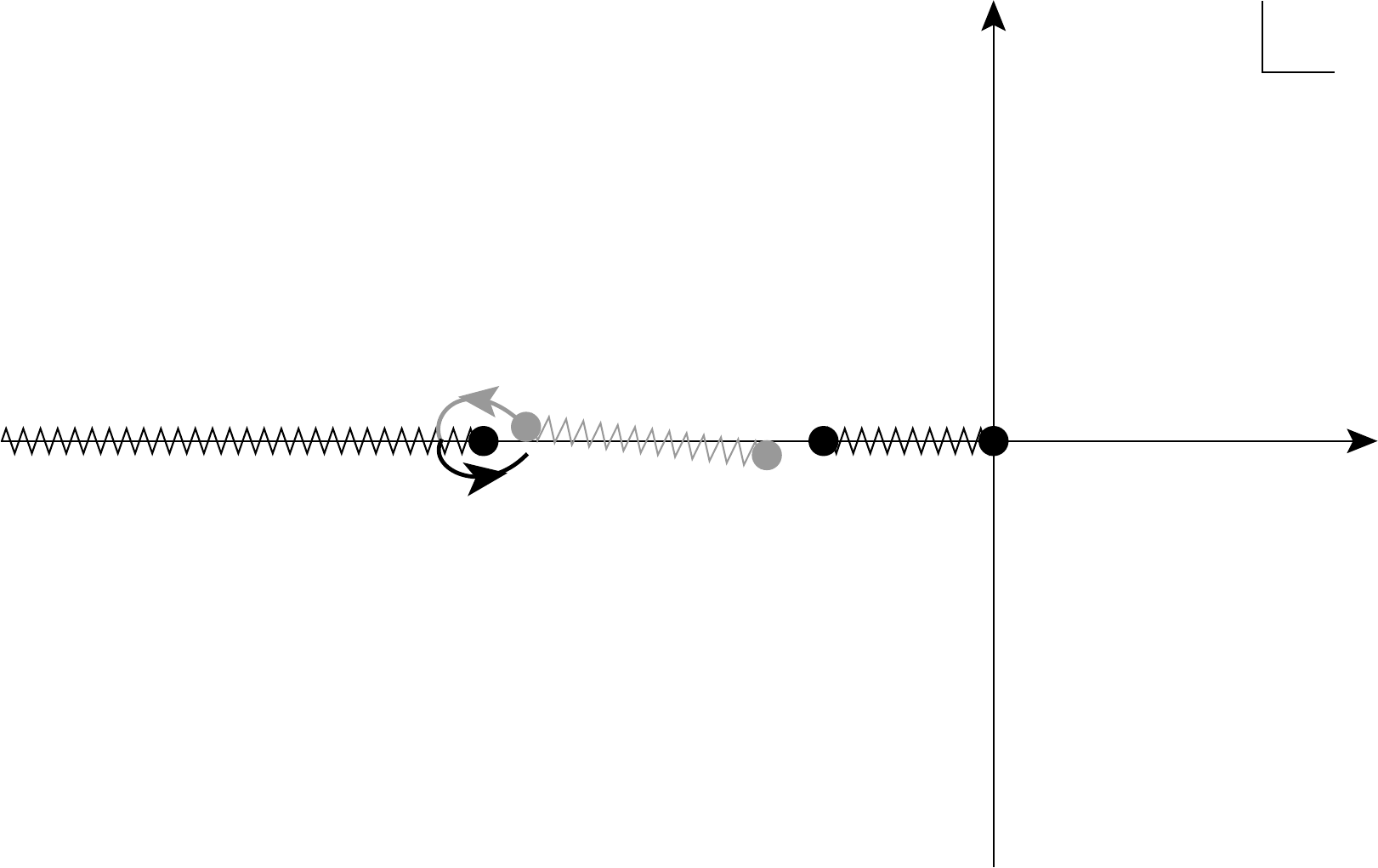}
	\end{picture}
	\caption{Sketch of the LHC structure of the resonance partial waves.}
	\label{fig:AnomalousThreshold}
\end{figure}

Let us consider the difference 
\beq
\Delta_\mathrm{cut}= s_+ - s_\mathrm{cut}^-
\eeq 
and add a small imaginary part to the virtualities, $q_i^2 \to q_i^2 + i \epsilon$. 
Then, for fixed values of $q_1^2$ we trace the path of $\Delta_\mathrm{cut}$ as a function of $q_2^2$. We find that
\begin{enumerate}
 \item $q_2^2 > \frac{(M_V^2-M_\pi^2)^2}{q_1^2}$: $\Delta_\mathrm{cut}$ has a positive real part and a small positive imaginary part of order $\epsilon$.
 \item $q_2^2 = \frac{(M_V^2-M_\pi^2)^2}{q_1^2}$: the imaginary part vanishes and the real part is negative (of order $\epsilon^2$).
 \item $q_2^2 < \frac{(M_V^2-M_\pi^2)^2}{q_1^2}$: the real part becomes again positive, the imaginary part is negative. 
\end{enumerate}
This implies that the square-root singularity of the discontinuity of the partial waves, which for $q_1^2 q_2^2 < (M_V^2-M_\pi^2)^2$ lies on the second sheet of the logarithmic LHCs, 
moves onto the physical sheet for $q_1^2 q_2^2 > (M_V^2-M_\pi^2)^2$, see the sketch in Fig.~\ref{fig:AnomalousThreshold}. 
This requires a deformation of the left-hand integration contour. In the case $q_1^2 q_2^2 < (M_V^2-M_\pi^2)^2$, the left-hand integral consists of two integrals
\begin{align}
	h^V(s) = \frac{1}{\pi} \int_{-\infty}^{s_\mathrm{cut}^-} ds' \frac{\Im h^V(s')}{s'-s} + \frac{1}{\pi} \int_{s_\mathrm{cut}^+}^0 ds' \frac{\Im h^V(s')}{s'-s},
\end{align}
while for $q_1^2 q_2^2 > (M_V^2-M_\pi^2)^2$ the dispersion integral picks up an anomalous contribution
\begin{align}
\label{hV_spacelike_anomalous}
	h^V(s) = \frac{1}{\pi} \int_{-\infty}^{s_\mathrm{cut}^-} ds' \frac{\Im h^V(s')}{s'-s} + \frac{1}{\pi} \int_{s_\mathrm{cut}^-}^{s_+} ds' \frac{\Delta_\mathrm{anom} h^V(s')}{s'-s} + \frac{1}{\pi} \int_{s_\mathrm{cut}^+}^0 ds' \frac{\Im h^V(s')}{s'-s}.
\end{align}
In the case of the Legendre functions, the normal imaginary part is given by
\begin{align}
	\Im\bigg( \frac{1}{\sigma_\pi(s) \lambda_{12}^{1/2}(s)}  Q_J(x_V) \bigg) = \frac{1}{\sigma_\pi(s) \lambda_{12}^{1/2}(s)} \frac{\pi}{2} P_J(x_V) \theta(1 - x_V^2).
\end{align}
Since the anomalous singularity is a square-root branch cut, the anomalous discontinuity is simply
\begin{align}
	\Delta_\mathrm{anom}\bigg( \frac{1}{\sigma_\pi(s) \lambda_{12}^{1/2}(s)}  Q_J(x_V) \bigg) = 2 \times \frac{1}{\sigma_\pi(s) \lambda_{12}^{1/2}(s)} \frac{\pi}{2} P_J(x_V),
\end{align}
which again can be verified by considering the path deformation in the complex $t$-plane.

The representation~\eqref{hV_spacelike_anomalous} indeed displays the correct integration regions, including anomalous contributions, but does not yet fully cover 
the realistic case encountered in $\gamma^*\gamma^*\to\pi\pi$, in which the singularities are stronger than in the schematic example discussed above.
In the case of higher partial waves, the discontinuity at the anomalous singularity behaves as $( s - s_+ )^{-(J+1)/2}$ due to $Q_J(x_V)$ alone. 
However, additional kinematic prefactors appear both in the partial waves and in the kernel functions, so that
in the realistic cases the anomalous singularity actually scales as $( s - s_+ )^{-5/2}$ for the $S$-waves, 
as $( s - s_+ )^{-7/2}$ for $\check h_{2,1}$ and $\check h_{45}^\pm$, 
and as $( s - s_+ )^{-9/2}$ for $\check h_{23}^\pm$.
Clearly, this is not integrable and the above representation~\eqref{hV_spacelike_anomalous} has to be modified further. 

To resolve this apparent contradiction, the important observation is that the contour integral around the anomalous singularity gives a non-vanishing contribution, so that the full anomalous integral is finite. The total anomalous integral can then be calculated as follows. We write
\begin{align}
	\label{eq:AnomalousDivergence}
	\frac{\Delta_\mathrm{anom} h^V(s')}{s'-s} = \sum_{k=0}^4 \frac{a_k(s)}{(s_+-s')^{(2k+1)/2}} + \tilde \Delta(s,s'),
\end{align}
where the first term collects the singular pieces of the integrand and $\tilde\Delta(s,s')$ vanishes as a square root for $s'\to s_+$. The coefficients $a_k(s)$ can be calculated analytically. 
The anomalous integral splits into two pieces
\begin{align}
	\label{eq:AnomalousIntegral}
	\frac{1}{2\pi i} \int_{\gamma_\mathrm{anom}} \hspace{-0.5cm} ds' \, \frac{h^V(s')}{s'-s} &= \frac{1}{\pi} \int_{s_\mathrm{cut}^-}^{s_+} ds' \tilde \Delta(s,s') - \frac{1}{\pi} \sum_{k=0}^4 \frac{2}{2k-1} \frac{a_k(s)}{(s_+-s_\mathrm{cut}^-)^{(2k-1)/2}}.
\end{align}
The integral around the singularity cancels exactly the singular pieces of the integral along the real axis, as can be seen by splitting the path into an integral up to $s_+ - \epsilon$ and a circular integral around the singularity with radius $\epsilon$. 

Finally, if the integral over $\tilde\Delta(s,s')$ is calculated numerically, one faces the problem of numerical instabilities close to $s_+$, given that $\tilde\Delta(s,s')$ is defined by the difference of two divergent expressions. This numerical issue can be handled by replacing $\tilde\Delta(s,s')$ close to $s_+$ by a fit function that has the same square-root-like behavior, i.e.
\begin{align}
	\sqrt{s_+ - s'} \sum_{k=0}^n b_k(s) {s'}^k
\end{align}
with some appropriate power $n$. The coefficients $b_k(s)$ are determined by a fit to $\tilde\Delta(s,s')$ in the vicinity of $s'=s_+$, but outside the region where numerical instabilities occur. 
The size of this region depends on the values of the virtualities, so that the fit region needs to be adapted accordingly.

We first verified that with this strategy we can indeed recover the original resonance partial waves from a representation such as~\eqref{hV_spacelike_anomalous}, even for 
large space-like virtualities that exceed the critical point~\eqref{virtualities_critical_point}. 
The generalization to the unitarized case with Omn\`es functions as in~\eqref{MO_solution_Swaves_Bachir} proceeds along the same lines, given that the Omn\`es functions do not 
alter the singularity structure, see App.~\ref{app:AnomalousThresholdTreatment} for more details. In this case, however, the derivatives of the Omn\`es function need to be provided as well, which in a numerically stable 
way follow from the spectral representation
\beq
\Omega_J^{(n)}(s)=\frac{n!}{\pi}\int_{4\mpi^2}^\infty ds'\frac{\Im\Omega_J(s')}{(s'-s)^{n+1}},
\eeq
or directly by taking derivatives of~\eqref{Omnes_def}. With increasing degree of singularity, numerical stability of the extrapolation becomes more of an issue, but even for the $-9/2$ singularities 
of $\check h_{23}^\pm$ remains under good control as long as the fit region is chosen prudently. 
However, we stress that for all $D$-waves besides $\check h_{45}^\pm$ the standard MO representation still applies, which does not involve integrals over the LHC. We verified that for $\check h_{2,1}$
and $\check h_{23}^\pm$ the above recipe for the treatment of the anomalous threshold in the modified MO representation indeed reproduces the same result as the standard MO representation.

	% !TEX root = ../DV_Paper.tex

\section{Numerics}
\label{sec:numerics}

In this section we present some numerical applications of the formalism developed in Sect.~\ref{sec:MO_solution}, mainly focused on 
the contribution of the $f_2(1270)$ resonance to the various helicity amplitudes. Experimentally, there is ample information
on the on-shell cross section $\gamma\gamma\to\pi^+\pi^-,\pi^0\pi^0$, derived from $e^+e^-\to e^+e^-\pi\pi$ 
via suitable cuts on the lepton momenta. 

\subsection{On-shell case}

In the on-shell case only the helicity amplitudes $H_{++}$ and $H_{+-}$ contribute. Adjusting the flux factor to an actual $\gamma\gamma$ initial state, one has
\begin{align}
\label{cross_section_onshell}
\frac{d\sigma}{d\Omega}\big(\gamma\gamma\to\pi^+\pi^-\big)&=\frac{\sigma_\pi(s)\alpha^2}{8s}\Big(\big|\bar H_{++}^\text{c}\big|^2+\big|\bar H_{+-}^\text{c}\big|^2\Big),\notag\\
\frac{d\sigma}{d\Omega}\big(\gamma\gamma\to\pi^0\pi^0\big)&=\frac{\sigma_\pi(s)\alpha^2}{16s}\Big(\big|\bar H_{++}^\text{n}\big|^2+\big|\bar H_{+-}^\text{n}\big|^2\Big),
\end{align}
where the particle-basis amplitudes are related to the isospin ones by the rotation given in~\eqref{isospin_matrix}.

To illustrate the behavior of the $f_2(1270)$, an isospin-$0$ $D$-wave resonance, we neglect unitarity corrections in the isospin-$2$ partial waves and combine our results for the $D$-waves 
with the $S$-waves from~\cite{Colangelo:2017qdm,Colangelo:2017fiz} (as well as the higher partial waves for the pion pole without rescattering). The only free parameters are then the photon couplings of the vector resonances $C_V$, which in a narrow-width picture are related to the
partial widths by means of~\eqref{CV}.
We find that the physical couplings do not exactly reproduce the cross section. 
This observation corresponds to the fact that the sum rules for the subtraction constants introduced in~\cite{GarciaMartin:2010cw} are not fulfilled exactly, pointing to a small correction
from higher intermediate states not explicitly included in the calculation.\footnote{A similar observation was made in~\cite{Danilkin:2018qfn}, where the authors argued that the difference between the fit values for the photon couplings and the ones extracted from the radiative widths reflected $SU(3)$ uncertainties. We disagree with that statement: if the deficit were due to $SU(3)$ uncertainties, it should
disappear once the known couplings for the individual states, $\omega$, $\rho^\pm$, $\rho^0$, are used instead of a common $SU(3)$ coupling, but this is not the case.}
To ensure agreement with the measured cross section, we therefore allow the couplings to vary, as a means to include phenomenologically the effect of higher intermediate states.

\begin{figure}[t]
	\centering
	\includegraphics[width=0.49\linewidth]{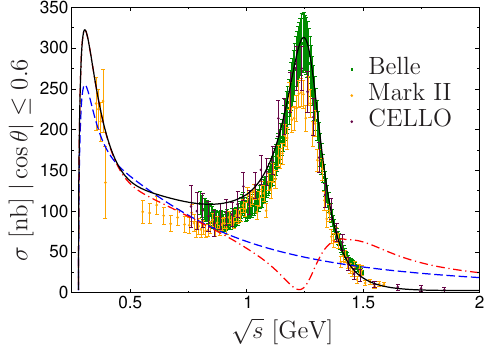}
	\includegraphics[width=0.49\linewidth]{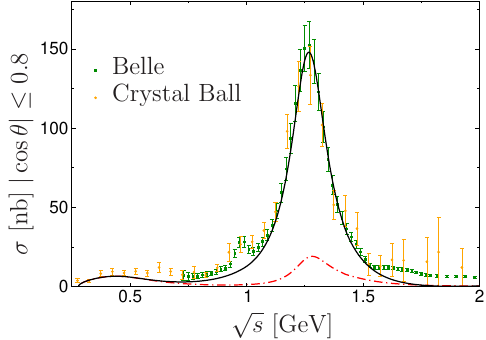}
	\caption{Cross section for $\gamma\gamma\to\pi^+\pi^-$ (left) and $\gamma\gamma\to\pi^0\pi^0$ (right), in comparison to the data from Belle~\cite{Mori:2007bu,Uehara:2009cka}, Mark II~\cite{Boyer:1990vu}, CELLO~\cite{Behrend:1992hy}, and Crystal Ball~\cite{Marsiske:1990hx}. The lines indicate the pion Born terms (blue dashed, all partial waves), including the $I=0$ unitarization of $S$- and $D$-waves (red dot-dashed), and the full solution (black solid).}
	\label{fig:onshell}
\end{figure}

Note that the experimental cross sections are not integrated over the full angular range, with $|\cos\theta|\leq 0.6$ and $|\cos\theta|\leq 0.8$ for the charged and neutral channels, respectively. The results in Fig.~\ref{fig:onshell} follow this convention. The relevant helicity amplitudes in the on-shell case are
\beq
h_{0,++}(s)=\frac{1}{2}\check h_{15}^+(s),\qquad
h_{2,++}(s)=\frac{s(s-4\mpi^2)}{2}\check h_{23}^+(s),\qquad
h_{2,+-}(s)=\frac{s-4\mpi^2}{2}\check h_{45}^+(s).
\eeq
In the figure, the blue dashed lines indicate the pion Born terms and the red dot-dashed ones their unitarization. The $S$-waves are treated as in~\cite{Colangelo:2017qdm,Colangelo:2017fiz}, with a phase shift from the inverse-amplitude method as specified in~\cite{Pelaez:2010fj}. This phase shift agrees well with dispersive $\pi\pi$ phase shift analyses~\cite{Colangelo:2001df,GarciaMartin:2011cn,Caprini:2011ky} at low-energies, but removes the $f_0(980)$ contribution in a controlled manner, which otherwise would require a coupled-channel treatment of the $\pi\pi/\bar K K$ $S$-waves.
Further, we do not include the $S$-waves resulting from the vector-meson exchanges, given that these contributions are not relevant for the $f_2(1270)$ and need to be studied together with the 
pion polarizabilities to ensure the correct low-energy properties of the $\gamma\gamma\to\pi\pi$ reaction, see~\cite{Colangelo:2017qdm,Colangelo:2017fiz}. 
These details can be improved most conveniently by introducing subtractions in the $S$-wave dispersion relations, but instead 
we focus here on the $f_2(1270)$ resonance, as it emerges mainly from the unitarization of the vector-meson $D$-waves, see Fig.~\ref{fig:onshell}, using the phase shift from~\cite{GarciaMartin:2011cn}. 
In the neutral channel, the unitarization of the Born terms alone actually results in a small resonant contribution, while in the charged channel it displays the pathological
behavior of the standard MO solution illustrated in~\eqref{MO_standard_zero}. In both cases the physical couplings need to be reduced by about $13\%$ to match the physical cross section, reflecting the 
impact of higher LHCs beyond the lightest vector mesons $\omega$, $\rho^\pm$, $\rho^0$ and potentially inelastic effects in the $\pi\pi$ $D$-wave.\footnote{The uncertainties for the radiative widths given in~\cite{Tanabashi:2018oca} are at the level of $3\%$ for $\omega\to\pi^0\gamma$ and $10\%$ for $\rho\to\pi\gamma$.}

\subsection{Singly- and doubly-virtual case}

Given that the $f_2(1270)$ resonance in the on-shell process can be largely understood as a unitarization of the vector mesons in the LHC, the only additional information required 
for the virtual processes concerns the $V\pi$ transition form factors as introduced in App.~\ref{app:Bornterms}, in analogy to the pion vector form factor for the pion-pole terms. 
For the $\omega$, this form factor is again available from a detailed dispersive analysis~\cite{Schneider:2012ez,Niecknig:2012sj,Danilkin:2014cra}. 
In contrast, a rigorous implementation of the $\rho$ should proceed in terms of $2\pi$ intermediate states, based on a suitable $\gamma^*\to3\pi$ amplitude~\cite{Hoferichter:2012pm,Hoferichter:2014vra,Hoferichter:2017ftn,Hoferichter:2018dmo,Hoferichter:2018kwz,Hoferichter:2019xyz}.
Here, we illustrate the numerical solution by approximating the dependence on the photon virtuality by a vector-meson-dominance (VMD) suppression $M_V^2/(M_V^2-q^2)$, which in the case of the pion
form factor reproduces the full solution very accurately~\cite{Colangelo:2017qdm,Colangelo:2017fiz,Colangelo:2018mtw}. For the $\omega$ transition form factor the deviations from VMD are more sizable, but a refined analysis
should address the $\rho$ LHC at the same time.

\begin{figure}[t]
	\centering
	\includegraphics[width=0.399\linewidth]{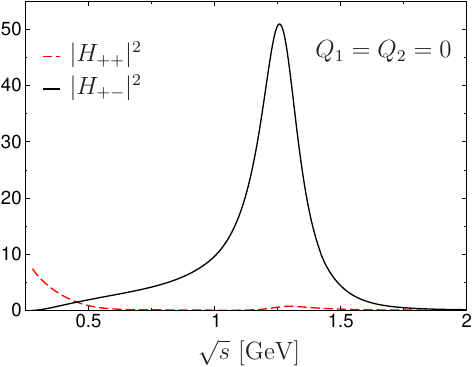}\quad
	\includegraphics[width=0.399\linewidth]{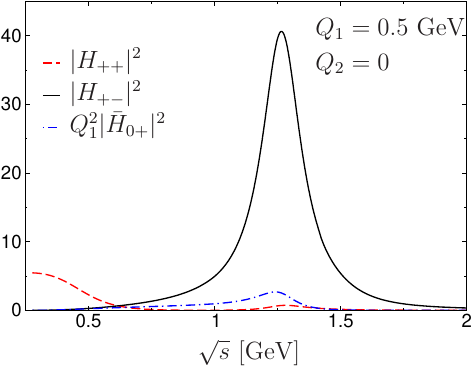}\\[-0.595cm]
	\includegraphics[width=0.399\linewidth]{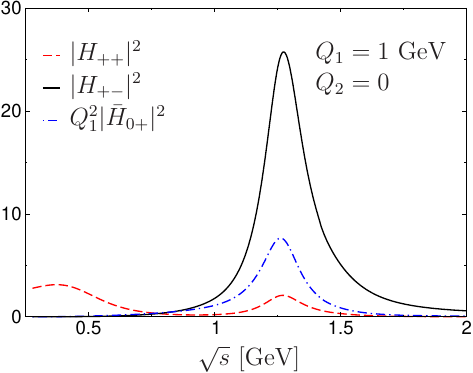}\quad
	\includegraphics[width=0.399\linewidth]{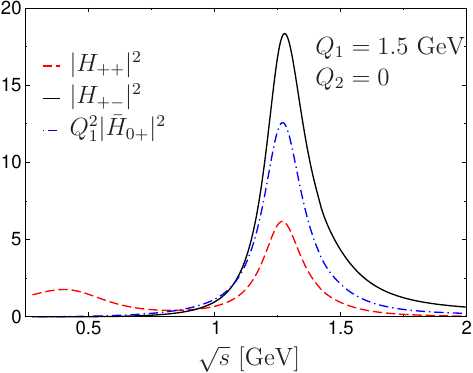}
	\caption{Angular-integrated helicity amplitudes~\eqref{hel_angular}, for singly-virtual virtualities $Q_i^2=-q_i^2$.}
	\label{fig:singly_virtual}
\end{figure}

For the virtual processes the canonical generalization of~\eqref{cross_section_onshell} would be
\begin{align}
\label{cross_section_offshell}
\frac{d\sigma}{d\Omega}\big(\gamma^*\gamma^*\to\pi^+\pi^+\big)&=\frac{\sigma_\pi(s)\alpha^2}{36\lambda_{12}^{1/2}(s)}
 \Big(2\big|\bar H_{++}^\text{c}\big|^2+2\big|\bar H_{+-}^\text{c}\big|^2-2q_2^2\big|\bar H_{+0}^\text{c}\big|^2-2q_1^2\big|\bar H_{0+}^\text{c}\big|^2+q_1^2q_2^2\big|\bar H_{00}^\text{c}\big|^2\Big),\notag\\
 \frac{d\sigma}{d\Omega}\big(\gamma^*\gamma^*\to\pi^0\pi^0\big)&=\frac{\sigma_\pi(s)\alpha^2}{72\lambda_{12}^{1/2}(s)}
 \Big(2\big|\bar H_{++}^\text{n}\big|^2+2\big|\bar H_{+-}^\text{n}\big|^2-2q_2^2\big|\bar H_{+0}^\text{n}\big|^2-2q_1^2\big|\bar H_{0+}^\text{n}\big|^2+q_1^2q_2^2\big|\bar H_{00}^\text{n}\big|^2\Big),
\end{align}
but we stress that these cross sections are not actual observables, only the $e^+e^-$ cross section is~\cite{Bonneau:1973kg,Budnev:1974de}, which can be seen from the fact that in the definition of~\eqref{cross_section_offshell} we needed to choose a convention for the flux factor and the counting of the polarization states (due to the latter this choice is discontinuous in the limit $q_i^2\to 0$). 
For these reasons we present our results instead directly in terms of the squared moduli of the helicity partial-wave amplitudes,
which are also the most relevant objects for the future application to HLbL scattering. For convenience, we combine $S$- and $D$-waves into
\beq
\label{hel_angular}
\frac{1}{2}\int^1_{-1} d\cos\theta\, |\bar H_{\lambda_1\lambda_2}|^2=\sum_J (2J+1)|h_{J,\lambda_1\lambda_2}|^2.
\eeq
Moreover, we focus on the $I=0$ amplitudes, where the unitarization effects that produce the $f_2(1270)$ occur.     

\begin{figure}[t]
	\centering
	\includegraphics[width=0.399\linewidth]{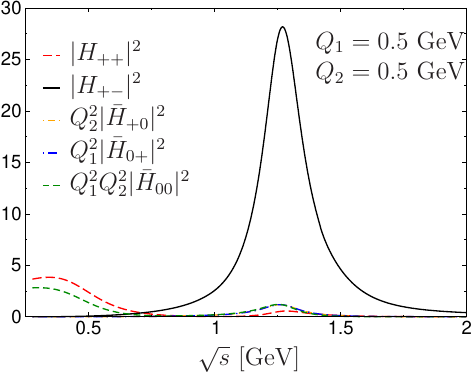}\quad
	\includegraphics[width=0.399\linewidth]{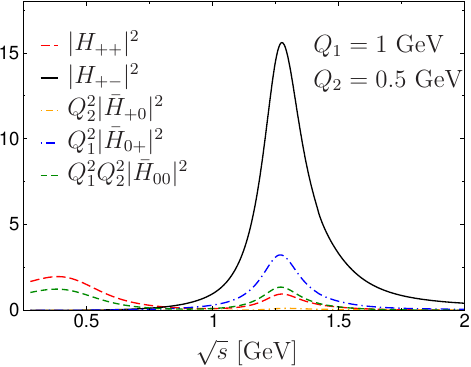}\\[-0.595cm]
	\includegraphics[width=0.399\linewidth]{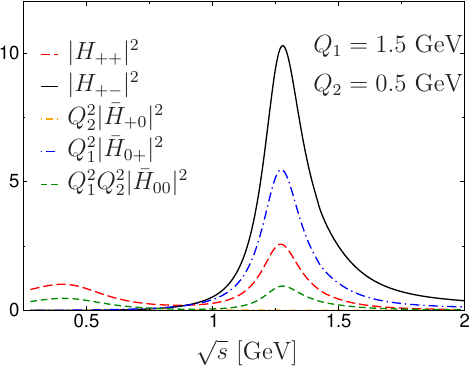}\quad
	\includegraphics[width=0.399\linewidth]{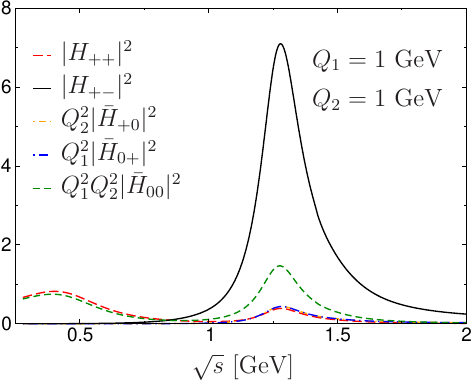}\\[-0.595cm]
	\includegraphics[width=0.399\linewidth]{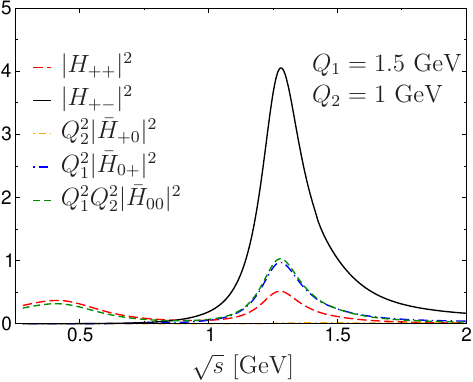}\quad
	\includegraphics[width=0.399\linewidth]{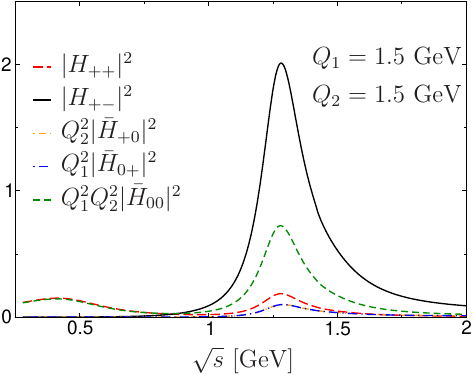}
	\caption{Angular-integrated helicity amplitudes~\eqref{hel_angular}, for doubly-virtual virtualities $Q_i^2=-q_i^2$.}
	\label{fig:doubly_virtual}
\end{figure}

The results are shown in Fig.~\ref{fig:singly_virtual} for several singly-virtual cases and in Fig.~\ref{fig:doubly_virtual} for doubly-virtual ones. Already for the on-shell case the Born terms appear 
suppressed compared to the $f_2(1270)$ peak, due to their enhancement in the cross section by the flux factor $1/s$, and that relative size does not change much once the virtualities are increased. 
In all cases, the $H_{+-}$ helicity amplitude gives the dominant effect, but the other helicity projections become increasingly important for larger virtualities. 
In addition, the overall size of the contribution decreases rapidly, as expected from the form factor suppression of the vector-meson couplings.

	% !TEX root = ../DV_Paper.tex

\section{Conclusions and outlook}
\label{sec:conlusions}

Dispersion relations for processes involving virtual photons require a careful study of 
the (helicity) amplitudes to ensure that results are not invalidated by kinematic singularities. 
A suitable such decomposition for the $\gamma^*\gamma^*\to\pi\pi$ amplitudes has been derived before
as a precursor to HLbL scattering~\cite{Colangelo:2015ama,Colangelo:2017qdm,Colangelo:2017fiz},
with first numerical solutions provided for the $S$-waves of the process. In this paper, 
we extended the solution to higher partial waves, introducing a new basis in which
that solution takes a simple form. In particular, we studied the role of vector mesons
in the left-hand cut of the amplitudes, in terms of which the $D$-wave resonance $f_2(1270)$ 
can be understood as an effect of the $\pi\pi$ final-state rescattering. 

Phenomenologically, the $D$-waves of $\gamma^*\gamma^*\to\pi\pi$ are indeed expected to contribute to HLbL scattering
mainly via the  $f_2(1270)$ resonance. For this application it is therefore crucial to understand all helicity amplitudes
of $\gamma^*\gamma^*\to\pi\pi$ including the potential role of subtraction constants. Here, we settled
this issue conclusively, detailing how the high-energy behavior of a given partial wave is tied to 
the necessity of subtractions in particular variants of the Muskhelishvili--Omn\`es solution, which explains 
why for the helicity amplitudes most relevant for the $f_2(1270)$ the standard variant fails, but a description
in terms of the left-hand singularities of the vector-meson amplitudes still applies. We then developed a strategy
how to cope with the anomalous thresholds that appear in the doubly-virtual case, even for space-like virtualities, 
and presented some numerical results for the helicity amplitudes that illustrate the role of the $f_2(1270)$ 
depending of the photon virtualities.

Our results will be crucial for a model-independent evaluation of the $f_2(1270)$ contribution to HLbL scattering in the 
anomalous magnetic moment of the muon, which so far has only been estimated within a Lagrangian-based hadronic model as a narrow resonance~\cite{Pauk:2014rta,Danilkin:2016hnh}.  
To this end, we demonstrated how all helicity amplitudes can be derived numerically 
from the unitarization of pion-pole and vector-meson-exchange contributions, with parameters determined 
from the comparison to the measured $\gamma\gamma\to\pi\pi$ cross section. The same intermediate states 
in the dispersion relation for HLbL scattering should thus allow one to capture effects corresponding to the $f_2(1270)$ 
beyond the narrow-width approximation, and without further assumptions on the form factors corresponding to helicity amplitudes 
that cannot be probed by available data. 
Even if data for the offshell process $\gamma^*\gamma^*\to\pi\pi$ were available,
currently under study at BESIII~\cite{Redmer:2018gah} and potentially in the future at Belle II~\cite{Kou:2018nap},
the weighting with respect to 
energies and virtualities in the $g-2$ integral need not resemble the one in the cross section, which makes 
a detailed understanding of the various helicity amplitudes all the more important.
In this way, the $f_2(1270)$ will be an important test case also for other resonances in the $1$--$2\GeV$ region that are hard to describe
explicitly in terms of their decay channels, but still need to be reliably estimated 
to confront the Standard-Model prediction for the muon $g-2$ at the level of accuracy anticipated
for the E989 Fermilab experiment~\cite{Grange:2015fou}.

\section*{Acknowledgements}
\addcontentsline{toc}{section}{Acknowledgements}

We thank G.~Colangelo for useful discussions and comments on the manuscript.
Financial support by the DOE (Grants No.\ DE-FG02-00ER41132
and DE-SC0009919) is gratefully acknowledged.

\begin{appendices}

% !TEX root = ../DV_Paper.tex

\section{Pion pole and resonance exchange}
\label{app:Bornterms}

The pion-pole contribution to the scalar functions $A_i$ reads~\cite{Colangelo:2015ama}
\begin{align}
\label{Ai_Born}
	\begin{split}
		A_1^\pi &=- F_\pi^V(q_1^2) F_\pi^V(q_2^2) \left( \frac{1}{t-M_\pi^2} + \frac{1}{u-M_\pi^2}\right), \\
		A_4^\pi &=- F_\pi^V(q_1^2) F_\pi^V(q_2^2) \frac{2}{s - q_1^2 - q_2^2} \left( \frac{1}{t-M_\pi^2} + \frac{1}{u-M_\pi^2}\right), \\
		A_2^\pi &= A_3^\pi = A_5^\pi = 0,
	\end{split}
\end{align}
where $F_\pi^V(q^2)$ refers to the electromagnetic form factor of the pion. As shown in~\cite{Colangelo:2015ama}, these expressions are identical
to the result in scalar QED multiplied by $F_\pi^V(q^2_i)$ to account for the photon virtualities. Accordingly, we use ``Born terms'' and ``pion pole''
interchangeably. Moreover, the overall sign is determined as for the Compton scattering process, because this sign 
does not depend on the conventions chosen for the pion field. For the partial-wave helicity amplitudes of $\gamma^*\gamma^*\to\pi\pi$
we then choose the sign in such a way that the helicity amplitudes in particle and isospin bases are related by
\beq
\label{isospin_matrix}
\begin{pmatrix}
H^\text{c}\\ H^\text{n} 
\end{pmatrix}
=\begin{pmatrix}
  \frac{1}{\sqrt{3}} & \frac{1}{\sqrt{6}}\\
  \frac{1}{\sqrt{3}} & -\sqrt{\frac{2}{3}}
 \end{pmatrix}
\begin{pmatrix}
H^0\\ H^2 
\end{pmatrix},\qquad
\begin{pmatrix}
H^0\\ H^2 
\end{pmatrix}
=\begin{pmatrix}
  \frac{2}{\sqrt{3}} & \frac{1}{\sqrt{3}}\\
  \sqrt{\frac{2}{3}} & -\sqrt{\frac{2}{3}}
 \end{pmatrix}
\begin{pmatrix}
H^\text{c}\\ H^\text{n} 
\end{pmatrix},
\eeq
for charged (c) and neutral (n) pions and isospin $I=0,2$, respectively, which in practice implies an overall sign in the transition from~\eqref{Ai_Born} to the helicity amplitudes.
In these conventions the Born-term partial-wave projections become~\cite{Colangelo:2017fiz}
\begin{align}
	\begin{split}
		N_{J,1}(s) &= F_\pi^V(q_1^2) F_\pi^V(q_2^2)  \left\{ \frac{8}{\sigma_\pi(s) \lambda_{12}^{1/2}(s)}\left(\frac{s q_1^2 q_2^2}{\lambda_{12}(s)} + M_\pi^2\right) Q_J(x) +2 \delta_{J0} 
		\frac{(q_1^2-q_2^2)^2-s (q_1^2+q_2^2)}{\lambda_{12}(s)} \right\} , \\
		N_{J,2}(s) &= F_\pi^V(q_1^2) F_\pi^V(q_2^2) \frac{2s \sigma_\pi(s)}{\lambda_{12}^{1/2}(s)} J \sqrt{\frac{(J-2)!}{(J+2)!}} \Big\{ 2x Q_{J-1}(x) - \left( (J+1)-x^2(J-1) \right) Q_J(x) \Big\} , \\
		N_{J,3}(s) &= F_\pi^V(q_1^2) F_\pi^V(q_2^2) \frac{4\sqrt{2s}\sigma_\pi(s)}{\lambda_{12}^{1/2}(s)} \frac{s}{s-q_1^2-q_2^2} \sqrt{\frac{J}{J+1}} x \Big\{ x Q_J(x) -Q_{J-1}(x) \Big\} , \\
		N_{J,4}(s) &= F_\pi^V(q_1^2) F_\pi^V(q_2^2) \frac{4\sqrt{2s}\sigma_\pi(s)}{\lambda_{12}^{1/2}(s)} \frac{q_1^2-q_2^2}{s-q_1^2-q_2^2} \sqrt{\frac{J}{J+1}} x \Big\{ x Q_J(x) - Q_{J-1}(x) \Big\} ,\\
		N_{J,5}(s) &= F_\pi^V(q_1^2) F_\pi^V(q_2^2) \frac{4}{\lambda_{12}(s)} \left\{ \frac{(q_1^2-q_2^2)^2-s^2}{\sigma_\pi(s) \lambda_{12}^{1/2}(s)} Q_J(x) + 2s \, \delta_{J0} \right\},
	\end{split}
\end{align}
with
\begin{align}
	x = \frac{s-q_1^2-q_2^2}{\sigma_\pi(s) \lambda_{12}^{1/2}(s)} 
\end{align}
and Legendre functions of the second kind
\begin{align}
	Q_J(x) = \frac{1}{2} \int_{-1}^1 \frac{P_J(z)}{x-z} dz.
\end{align}
In particular, the isospin matrices in~\eqref{isospin_matrix} ensure that the standard form of Watson's theorem~\cite{Watson:1954uc} holds, i.e.\ in the elastic regime 
the phases of the $\gamma^*\gamma^*\to\pi\pi$ helicity partial waves
agree with the corresponding $\pi\pi$ phase shifts.

In the same way, the exchange of a vector meson based on a Lagrangian model~\cite{GarciaMartin:2010cw} leads to
\begin{align}
\label{Ai_V}
	\begin{split}
		 A_1^V &= \frac{1}{2}C_V^2F_{V\pi}(q_1^2)F_{V\pi}(q_2^2)  \bigg( \frac{4t+4M_\pi^2-s-q_1^2-q_2^2}{t-M_V^2}+\frac{4u+4M_\pi^2-s-q_1^2-q_2^2}{u-M_V^2}\bigg), \\
		 A_2^V &= - A_4^V = C_V^2 F_{V\pi}(q_1^2)F_{V\pi}(q_2^2) \bigg( \frac{1}{t-M_V^2} + \frac{1}{u-M_V^2}\bigg), \\
		 A_3^V &= - C_V^2 F_{V\pi}(q_1^2)F_{V\pi}(q_2^2) \frac{1}{s - \Sigma_{\pi\pi}+2M_V^2} \bigg( \frac{1}{t-M_V^2} + \frac{1}{u-M_V^2}\bigg), \\
		 A_5^V &= 0,
	\end{split}
\end{align}
where $F_{V\pi}(q^2)$ denotes the $V\pi$ transition form factor and $C_V$ is the coupling in the Lagrangian model, related to the decay width $V\to\pi\gamma$ by~\cite{GarciaMartin:2010cw}
\begin{align}
\label{CV}
	\Gamma_{V\to\pi\gamma} = \frac{\alpha}{2} C_V^2 \frac{(M_V^2-M_\pi^2)^3}{3M_V^3}.
\end{align}
The helicity partial waves are
\begin{align}
		h_{J,1}^V(s) &= - C_V^2 F_{V\pi}(q_1^2) F_{V\pi}(q_2^2) \begin{aligned}[t]
							& \bigg\{ \frac{4}{\sigma_\pi(s) \lambda_{12}^{1/2}(s)} Q_J(x_V) f_1^V(s) \\
							& - \delta_{J0} \left[ 2s + \Big(s - \Sigma_{\pi\pi}+2M_V^2\Big)\frac{(q_1^2-q_2^2)^2 - s(q_1^2+q_2^2)}{\lambda_{12}(s)} \right] \bigg\}, \end{aligned} \notag\\
		h_{J,2}^V(s) &= C_V^2 F_{V\pi}(q_1^2) F_{V\pi}(q_2^2) \frac{s (s-q_1^2-q_2^2) \sigma_\pi(s)}{\lambda_{12}^{1/2}(s)} J \sqrt{\frac{(J-2)!}{(J+2)!}} \notag \\
			&\qquad \times \Big\{ 2x_V Q_{J-1}(x_V) - \left( (J+1)-x_V^2(J-1) \right) Q_J(x_V) \Big\}, \notag\\
		h_{J,3}^V(s) &= C_V^2 F_{V\pi}(q_1^2) F_{V\pi}(q_2^2) \bigg(s-\frac{\lambda_{12}(s)}{s-\Sigma_{\pi\pi} + 2M_V^2}\bigg) \frac{2\sqrt{2s}\sigma_\pi(s)}{\lambda_{12}^{1/2}(s)}  \sqrt{\frac{J}{J+1}} x_V \Big\{ x_V Q_J(x_V) -Q_{J-1}(x_V)\Big\}, \notag\\
		h_{J,4}^V(s) &= C_V^2 F_{V\pi}(q_1^2) F_{V\pi}(q_2^2)(q_1^2-q_2^2) \frac{2\sqrt{2s}\sigma_\pi(s)}{\lambda_{12}^{1/2}(s)} \sqrt{\frac{J}{J+1}} x_V \Big\{ x_V Q_J(x_V) - Q_{J-1}(x_V) \Big\},\notag\\
		h_{J,5}^V(s) &= - C_V^2 F_{V\pi}(q_1^2) F_{V\pi}(q_2^2) \bigg\{ \frac{16}{\sigma_\pi(s) \lambda_{12}^{1/2}(s)} Q_J(x_V) f_5^V(s) - 4s \delta_{J0} \frac{s - \Sigma_{\pi\pi}+2M_V^2}{\lambda_{12}(s)} \bigg\},
\end{align}
where
\begin{align}
	\label{eq:HelAmpAbbreviations}
	\begin{split}
		f_1^V(s) &= M_V^2 (s-q_1^2-q_2^2) +  \frac{\big((M_V^2-M_\pi^2)^2 + q_1^2 q_2^2 \big) \big( (q_1^2-q_2^2)^2-s(q_1^2+q_2^2) \big) - 4s q_1^2 q_2^2(M_V^2-M_\pi^2)}{\lambda_{12}(s)}, \\
		f_5^V(s) &= \frac{s(M_V^2-M_\pi^2)^2 + s\big(q_1^2 q_2^2 + M_V^2(s-q_1^2-q_2^2) \big) + M_\pi^2\big( (q_1^2-q_2^2)^2-s(q_1^2+q_2^2) \big)}{\lambda_{12}(s)}, \\
		x_V &= \frac{s - \Sigma_{\pi\pi}+2M_V^2}{\sigma_\pi(s) \lambda_{12}^{1/2}(s)}.
	\end{split}
\end{align}

% !TEX root = ../DV_Paper.tex

\section{Anomalous singularities in the modified MO representation}
\label{app:AnomalousThresholdTreatment}

In this appendix, we explain in more detail how the dispersive integrals over the anomalous LHC can be computed in a numerically stable way. The appearance of the Omn\`es function in the unitarized case leads to additional complications compared to the description in Sect.~\ref{sec:resonance_partial_waves}. Instead of~\eqref{eq:AnomalousIntegral}, the anomalous integral is given by
\begin{align}
	I &:= \frac{1}{2\pi i} \int_{\gamma_\text{anom}} ds' \frac{h^V(s')}{\Omega(s') (s'-s)} .
\end{align}
One could proceed as in~\eqref{eq:AnomalousIntegral} and directly subtract the part of the expanded integrand that diverges for $s' \to s_+$. However, in this case, the numerical instabilities in the function $\tilde\Delta(s,s')$ become worse, since the cancellation of the two divergent expressions involves the Omn\`es function and its derivatives, both of which are calculated only numerically. However, these intricate cancellations can be avoided if not the full integrand including the Omn\`es function is expanded but only the part involving the resonance LHC, exactly as in~\eqref{eq:AnomalousDivergence}. We define
\begin{align}
	g(s,s') &:= \sum_{k=0}^4 \frac{(-1)^{k+1} a_k(s)}{(s' - s_+)^{(2k+1)/2}} ,
\end{align}
so that
\begin{align}
	\Im g(s,s') &= \sum_{k=0}^4 \frac{a_k(s)}{(s_+ - s')^{(2k+1)/2}} , \quad s' < s_+ ,
\end{align}
which leads to
\begin{align}
	I &= \frac{1}{\pi} \int_{s_\mathrm{cut}^-}^{s_+} ds' \frac{\tilde \Delta(s,s')}{\Omega(s')} + \underbrace{ \frac{1}{2\pi i} \int_{\gamma_\text{anom}} ds' \frac{1}{\Omega(s')} g(s,s') }_{=: I_2}.
\end{align}
The first integral is manifestly finite and the cancellation of divergences in $\tilde \Delta(s,s')$ is identical to the case without unitarization, in particular no further instabilities are introduced by derivatives of the Omn\`es function. The second integral can be split into a path up to close to the singularity at $s_+$ and an integral circling around the singularity:
\begin{align}
	\label{eq:AnomalousIntegralSplit}
	I_2 &= \frac{1}{\pi} \int_{s_\mathrm{cut}^-}^{s_+ - \epsilon} ds' \frac{1}{\Omega(s')} \Im g(s,s') + \frac{\epsilon}{2\pi} \int_{\pi}^{-\pi} d\phi e^{i\phi} \frac{1}{\Omega(s'(\phi))} g(s,s'(\phi))  .
\end{align}
For $\epsilon\to0$, both integrals contain divergent pieces that cancel in the sum. The result can be obtained by multiple integration by parts, leading to
\begin{align}
	I_2 &= \frac{1}{\pi} \int_{s_\mathrm{cut}^-}^{s_+} ds' \frac{\sqrt{s_+-s'}}{\Omega(s')} \sum_{k=0}^4 a_k(s) b_k(s') + \frac{1}{\pi} \frac{1}{\Omega(s_\mathrm{cut}^-)} \sum_{k=0}^4 \frac{c_k(s,s_\mathrm{cut}^-)}{(s_+-s_\mathrm{cut}^-)^{(2k-1)/2}} .
\end{align}
The integral in the first term has to be done numerically but does not introduce any instabilities as the integrand vanishes as a square root for $s'\to s_+$. The functions $b_k(s')$ depend on the first $k+1$ derivatives of the Omn\`es function at $s'$. The second term denotes the lower boundary term of the integration by parts at $s_\mathrm{cut}^-$. The divergent upper boundary term at $s_+-\epsilon$ has canceled against the circular integral in~\eqref{eq:AnomalousIntegralSplit}. The above expression is numerically stable as long as $s_+$ is not too close to $s_\mathrm{cut}^-$, i.e.\ as long as we stay away from the exceptional point $q_1^2 q_2^2 = (M_V^2 - M_\pi^2)^2$. In the vicinity of this singular point, one can further expand the coefficients $a_k(s)$ around $q_1^2 \sim \frac{(M_V^2 - M_\pi^2)^2}{q_2^2}$ to obtain a manifestly finite expression.

\end{appendices}

\renewcommand\bibname{References}
\renewcommand{\bibfont}{\raggedright}
\bibliographystyle{utphysmod}
\phantomsection
\addcontentsline{toc}{section}{References}
\bibliography{Literature}

\end{document}